\documentclass[twocolumn,showpacs,preprintnumbers,amsmath,amssymb,epsf]{revtex4}
\usepackage{graphicx}% Include figure files
\usepackage{dcolumn}% Align table columns on decimal point
\usepackage{epsf}
\usepackage{bm}% bold math
\begin{document}

\title{Disorder-induced magnetic memory: Experiments and theories}

\author{M.~S.~Pierce$^{1,2}$, C.~R.~Buechler$^{1}$, L.~B.~Sorensen$^{1}$,
S.~D.~Kevan$^{3}$,
E.~A.~Jagla$^{4}$,
J.~M.~Deutsch$^{5}$,T.~Mai$^{5}$, O.~Narayan$^{5}$,
J.~E.~Davies$^{6}$, Kai~Liu$^{6}$, G.~T.~Zimanyi$^{6}$,
H.~G.~Katzgraber$^{7}$,
O.~Hellwig$^{8}$, E.~E.~Fullerton$^{8}$,
P.~Fischer$^{9}$, and J.~B.~Kortright$^{9}$.
}

\address{$^{1}$Department of Physics, University of Washington,  
Seattle, Washington 98195, USA}

\address{$^{2}$Materials Science Division, Argonne National Laboratory, Argonne Illinois 60439, USA}

\address{$^{3}$Department of Physics, University of Oregon, Eugene,  
Oregon 97403, USA}

\address{$^{4}$Centro At\'omico Bariloche, Comisi\'on Nacional de  
Energ\'{\i}a At\'omica, (8400) Bariloche, Argentina}

\address{$^{5}$Department of Physics, University of California, Santa  
Cruz, California 95064, USA}

\address{$^{6}$Department of Physics, University of California, Davis,  
California 95616, USA}

\address{$^{7}$Theoretische Physik, ETH Z\"urich, CH-8093  
Z\"urich, Switzerland}

\address{$^{8}$Hitachi Global Storage Technologies, San Jose,  
California 95120, USA}

\address{$^{9}$Lawrence Berkeley National Laboratory, Berkeley,  
California 94720, USA}

\date{\today}

\begin{abstract}

Beautiful theories of magnetic hysteresis
based on random microscopic disorder have been developed over the past ten years. Our
goal was to directly compare these theories with precise experiments.  To do so, 
we first developed and
then applied coherent x-ray speckle metrology to a series of thin multilayer perpendicular
magnetic materials.  To directly observe the effects of disorder, we deliberately introduced 
increasing degrees
of disorder into our films.  We used coherent x-rays, produced at the Advanced Light Source at Lawrence
Berkeley National Laboratory, to generate highly speckled magnetic scattering patterns.  The apparently
``random" arrangement of  the speckles is due to the exact configuration of the magnetic domains in the
sample. In effect, each speckle pattern acts as a unique fingerprint for the magnetic domain
configuration. Small changes in the domain structure change the speckles, and comparison of the
different speckle patterns provides a quantitative determination of how much the domain structure has
changed.  Our experiments quickly answered one longstanding question: How is the magnetic domain
configuration at one point on the major hysteresis loop related to the configurations at the same point
on the loop during subsequent cycles?  This is called microscopic return-point memory (RPM).
We found the RPM is partial and imperfect
in the disordered samples, and completely absent when the disorder 
was below a threshold level.
We also introduced and answered a second important new question: How
are the magnetic domains at one point on the major loop related to the domains at the complementary
point, the inversion symmetric point on the loop, during the same and during subsequent cycles?  This is
called microscopic complementary-point memory (CPM). We found the CPM is also partial and imperfect  in the disordered samples and completely absent when the disorder was not present.  In
addition, we found that the RPM is always a little larger than the CPM. 
We also studied the correlations between the domains within a single ascending or descending loop.
This is called microscopic half-loop memory (HLM) and enabled us to measure the degree of change in the domain structure due to changes in the applied field. 
No existing theory was capable
of reproducing our experimental results. So we developed new theoretical models
that do fit our experiments. Our experimental and theoretical results 
set new benchmarks for future work.
\end{abstract}


\maketitle

\section{Introduction
}

What causes magnetic hysteresis and how is it induced and influenced by coexisting
microscopic disorder?  This is the question that we address and provide new answers
about in this paper.  We are able to provide new information about this
venerable old question because we have developed a new way to directly
probe the effect of disorder on the spatial structure of the
microscopic magnetic domain configuration as a function of the applied
magnetic field history.  When we finished our experimental study, we
discovered that our results could not be explained by any existing
microscopic theories of magnetic hysteresis.  So we developed several viable theoretical models.  In this
paper, we present our detailed experimental results and the theoretical
models that we developed to explain them.

Magnetic hysteresis is fundamental to all magnetic storage technologies
and consequently is a cornerstone of the present information age.
The magnetic recording industry deliberately introduces carefully
controlled disorder into its materials to obtain the desired hysteretic
behavior and magnetic properties. Over the past 40 years, such magnetic hardening has developed
into a high art form. However, despite decades of intense study
and significant recent advances, we still do not have a
completely satisfactory microscopic understanding of magnetic
hysteresis.

The exponential growth of computing power that fueled the information age has been driven by two
technological revolutions: 1) The integrated circuit revolution and its exponential growth described by
Moore's Law.  2) The magnetic disk drive revolution and its exponential growth which for the past decade
has surpassed Moore's Law.  Both of these mature technologies are rapidly approaching their fundamental
physical limits.  If the incredible growth rate of storage capacity in magnetic media
is to continue, new advances in our fundamental understanding of magnetic hysteresis are needed.

For the past 20 years, magnetic films with perpendicular anisotropy 
have been extensively studied for
their potential to extend the limits of storage capacity.  
Early in 2005, the first
commercial disk drives using perpendicular magnetic media 
became available.  The system that we study here is a model 
for these new perpendicular magnetic media.
In this paper, we present our results on the effect of
disorder on the correlations between the domain configurations 
in these systems.

To study the detailed evolution of the magnetic domain configuration 
correlations in our samples, we developed a new x-ray
scattering technique, coherent x-ray speckle metrology (CXSM).  We illuminate our samples with coherent
x-rays tuned to excite virtual 2p to 3d resonant transitions in cobalt.  The resulting resonant excitation of
the cobalt provides our magnetic signal.  The coherence of the x-rays produces a magnetic
x-ray speckle pattern.  The positions and intensity of the speckles provides a detailed fingerprint of
the microscopic magnetic domain configuration.  Changes in the magnetic domain configuration
produce changes in the speckle pattern.  So by comparing these magnetic fingerprints versus the
magnetic field history---by cross-correlating speckle patterns with different magnetic field
histories---we obtain a quantitative measure of the applied field-history-induced evolution of the magnetic
domain configuration.

Here we report our results obtained by applying CXSM to investigate the
effects of controlled disorder on the magnetic domain evolution in a
series of Co/Pt multilayer samples with perpendicular anisotropy.  We introduced disorder into the samples by systematically increasing the interfacial
roughness of the Co/Pt multilayers during the growth process.  We found
that this disorder induces memory in the microscopic magnetic domain
configurations from one cycle of the hysteresis loop to the next, despite taking the samples
through magnetic saturation.  Our
lowest disorder samples have no detectable cycle-to-cycle memory; their
domain patterns are unique each time the sample is cycled around the
major loop.   As we increase the disorder, the cycle-to-cycle memory
develops and grows to a maximum value, but never becomes perfect or complete at room
temperature.

In this paper, we only present our results for microscopic magnetic
memory along the major loop in the slow field sweep limit.
In this limit, the measured hysteresis loop is the same over many 
decades of sweep rate.  The hysteresis in this limit is often
called rate-independent hysteresis, or quasi-static hysteresis.
There are, of course, also interesting and important hysteresis
effects that occur at high sweep rates.  Our strategy was to study
the simpler rate-independent hysteresis case before adding the
additional physics, and complications, associated with high sweep
rates.  As we explain below, the disorder dependence of
the rate-independent hysteresis in our system turned 
out to be remarkably rich and interesting.
We do not discuss our results for rate-independent minor 
loop memory in this paper, but we briefly reported them 
recently \cite{prl1}.

The best modern microscopic disorder-based theories of magnetic hysteresis
were built on the foundations of Barkhausen noise measurements \cite{sethna-dahmen}.
Even in the
rate-independent limit, the magnetization of a disordered ferromagnet
does not change smoothly as the applied field is swept up and down.
Instead, there are magnetic domain avalanches that produce
magnetization jumps.  These avalanches exhibit power-law
size distributions
indicating that many different size regions change their magnetization
in jumps as the field is swept around the major hysteresis loop.

A
comprehensive, recent review of Barkhausen noise studies---including a
translation of Barkhausen's 1919 paper---is given in Ref. \cite{zapperi}.
For some materials in the rate-independent limit, the Barkhausen noise
is independent of the magnetic sweep rate; these avalanches occur at
fixed values of the applied field, independent of the sweep rate \cite{zapperi_prl}.

Barkhausen measurements provide exquisite information about the time
structure of the avalanches, but they usually do not provide any spatial
information about the location of the avalanches.
Because we directly measure the nanometer scale spatial structure of
the magnetic domain configuration changes, we obtain detailed information about the configuration evolution
that cannot be obtained directly from
the best classical Barkhausen noise studies or
from their modern optical implementations \cite{optical-bark}.
Because there has been extensive theoretical work on Barkhausen noise,
the corresponding field-history-dependent microscopic morphologies
of the magnetic domain
configurations have been indirectly inferred from the Barkhausen
time-signals via detailed computer simulations.  For example, Sethna,
Dahmen and their co-workers have shown that the morphology for
their random field Ising model (RFIM) is fractal in space.
They provide a comprehensive review of their work
in Ref. \cite{sethna-dahmen}.

Taken together, the detailed fractal-in-time structure measured via the
Barkhausen noise, and the extensive computer simulations by Sethna,
Dahmen, and others, imply that their magnetic domain configurations are
fractal in space.  Therefore, why not simply measure the correlations between the magnetic domain
configurations directly?  That is precisely what we do in this paper.
There has been very little systematic, ensemble-level experimental work on the spatial evolution
of the magnetic domain configurations \cite{optical-bark, waveform_magnetics}, but this information is readily
available from the existing simulations.  However, up until now almost
all of the work has been done for pure RFIMs.  Our experimental system
and the new generation of perpendicular magnetic disk drive media have
long-range dipole interactions.  This means that new theories that
include the dipolar interactions\cite{dipolar_rmp} will be required to understand these materials.

During our work, we unearthed three interesting aspects of our magnetic
domain wall evolution.  The first, called major loop return-point
memory (RPM), describes the magnetization for each point on the major
loop.  If this magnetization is precisely the
same for each cycle around the major loop, then we have macroscopic
major loop RPM.
If, in addition, the microscopic magnetic domain configuration is also
identical, then we have microscopic major loop RPM.  Our
experiments show that our samples have perfect macroscopic
major loop RPM, but imperfect microscopic major loop RPM
at room temperature.  

The second,
called complementary-point memory (CPM), describes the inversion
symmetry of the major loop through the origin.  If the
magnetization at field $H$ on the descending branch is equal to the
minus the magnetization at field $-H$ on the ascending branch,
then we have perfect macroscopic major loop CPM.
If, in addition, the magnetic domains are precisely reversed,
then we have perfect microscopic major loop CPM.  Our experiments
show that our samples have perfect macroscopic major loop CPM, 
but imperfect microscopic major loop CPM at room temperature.
In addition, we find that our measured values for the 
microscopic RPM are consistently a little larger than
those for our microscopic CPM---thus the RPM-CPM symmetry
is slightly broken.  

The third, called half-loop-memory(HLM), describes the degree of change in the magnetic domain configurations along a single branch of the major hysteresis loop.  Our experiments show that disorder has a direct effect on how the domains evolve.  The greater the disorder present in the sample, then the greater the observed changes in the domain configurations as the applied field is slowly adjusted to take the system along the major hysteresis loop.  Our measured values for the HLM are consistently higher in the low disorder samples than those present in the disordered samples.

We were inspired to do our experimental study by the beautiful work on
the RFIM by Sethna, Dahmen, and coworkers \cite{sethna-dahmen}.  We were therefore very
surprised to discover that their model could not describe our
experimental results.  Their pure zero-temperature RFIM predicts
perfect macroscopic and microscopic major loop RPM,
but it does not agree with our experiments because it predicts
essentially no microscopic major loop CPM.
It seems reasonable that their $T>0$
RFIM will predict perfect macroscopic RPM
but imperfect microscopic RPM
like that observed in our experiments,
but this has not been tested.  However their model cannot
predict our observed microscopic CPM and therefore it also cannot 
predict the slightly broken microscopic RPM-CPM symmetry that our experiments observe.

So, what physics is required to produce imperfect microscopic RPM 
and CPM with the slightly broken symmetry?  There are two aspects
to this question---the imperfection and the RPM-CPM symmetry breaking.
Almost all models have perfect memory at $T=0$ and imperfect memory 
for $T>0$.  And it seems likely that the imperfect memory that we
observe could be caused by temperature effects, but this has 
not yet been established.  On the other hand, no viable theoretical 
model for the slight RPM-CPM symmetry breaking existed.  So we
developed viable models.  The key idea behind each of our models
was to combine physics with
spin-reversal symmetry with physics without spin-reversal symmetry.
Then the spin-reversal-symmetric physics produces symmetric memory
${{\rm RPM} = {\rm CPM}}$ and the non-symmetric physics produces 
symmetry-broken memory ${{\rm RPM} \not= {\rm CPM}}$.

Within the standard RXIM models---{\it {viz.}}, RAIM, RBIM, and 
RCIM, and RFIM where A denotes anisotropy, B denotes bond, C denotes
coercivity, and F denotes field---the first three have
spin-reversal symmetry, but the fourth (RFIM) does not.  
So one way to produce slightly symmetry-broken memory
is to combine the RFIM with one of the symmetric models.
Surprisingly, another way is to combine one of the
symmetric models with vector spin dynamics because vector
dynamics breaks the spin-reversal symmetry.  We report
our work on three viable models:
Model 1 combines a pure RFIM with a pure RCIM.
Model 2 combines a pure RAIM with vector spin dynamics.
Model 3 combines a pure RFIM with a pure spin-glass model

We explored Model 1 and Model 2 in the most detail.
By tuning the model parameters, we were able to 
semi-quantitatively match our
experimentally observed disorder-dependence
and magnetic-field-dependence of
(i) the domain configurations,
(ii) the shape of the major loops,
(iii) the values of the RPM and CPM, and
(iv) the slight RPM-CPM symmetry breaking.

Note that in order to properly describe our observed magnetic domain
configurations, we had to include the long-range dipolar
interactions. In contrast to the Sethna-Dahmen 
RFIMs that predict spatially fractal magnetic domain
configurations \cite{sethna-dahmen},
our samples exhibit labyrinthine domain configurations
due to their long-range dipolar interactions.

The remainder of this paper is organized as follows.
Section 2 describes the physics of return point memory and
complementary point memory.
Section 3 describes our experiments, sample fabrication,
structural characterization, magnetic characterization, and
coherent x-ray speckle metrology (CXSM) characterization.
Section 4 describes our data analysis methodology.
Section 5 describes the results of our data analysis.
Section 6 describes the theoretical models that we developed
to account for the observed behavior of our system.
Section 7 presents our conclusions.

\section{Macroscopic and Microscopic Return-Point Memory
and Complementary Point Memory
}

In his 1903 dissertation at G\"ottingen entitled ``On the magnetization
produced by fast currents and the operation of Rutherford-Marconi
magnetodetectors," Erwin Madelung presented his rules for
magnetic hysteresis as illustrated in Figure 1:

\begin{enumerate}

\item  Major-loop return-point memory.

The magnetization of the sample at every point on the major loop is 
completely
determined only by the applied field, and all first-order reversal 
curves starting from the major loop and going to saturation are 
uniquely determined by their starting point.
The curve $1\rightarrow +S$ in Fig. 1 illustrates a first-order 
reversal curve (forc).

\item  Minor-loop return-point memory.

The magnetization of the sample at every point on the major loop is 
completely determined solely by the value of the applied field, even
when the point on the
major loop is reached starting from a point inside the major loop. This 
holds
for every order reversal curve.
The curve $2\rightarrow1$ illustrates this property for a second-order
reversal curve (sorc).

\item The memory deletion property, a.k.a. the wiping out property.

The magnetization of the sample at every point on a reversal
curve is precisely the same as that for its parent curve as soon
the reversal curve returns to its parent.
In this way, all memory of the previous field history between
the initial departure from the parent and the return to the parent
has been erased. This holds for every order reversal curve.
The curve $3 \rightarrow +S$ illustrates this for a third-order
reversal curve (torc).

\item The congruency property.

All return curves that start from reversal at the same value of the 
applied
field have the same shape thereafter independent of the entire previous
applied field history.

\item The similarity property for initial magnetization curves.

When any initial magnetization curve is reversed at point a, the
reversed return curve to saturation will pass through the inversion 
symmetric point to a as it proceeds to saturation.
As discussed below, we call the analogous property to the similarity 
property---for reversal curves that do not start from a point on the
initial magnetization curve---the complementary-point memory property.

\end{enumerate}

Madelung formulated his rules based on his careful experimental
studies of different alloys of steel and published them
in 1905 and 1912 \cite{madelung_05}.
Because Madelung formulated his rules before the existence
of magnetic domains was known, he only considered the macroscopic 
magnetization.  Nevertheless, his rules still predict the
macroscopic magnetization of ``any typical" sample versus its
applied field history.  Madelung's rules have truly been the foundation
for all modern theories of hysteresis.

It is therefore surprising that Madelung's rules are so rarely cited.
Apparently this is because essentially all of the subsequent work
has been focused on the Preisach model. The obscurity of
Madelung's magnetic hysteresis work is particularly surprising
because the Preisach model has been well known to be unphysical
for a very long time due to heavy reliance on phenomenology.

Of course, Madelung's rules do not apply to every magnetic
system.  For example, many systems exhibit accomodation,
reptation and magnetic viscosity effects, and all systems
exhibit dynamic hysteresis effects. However, on the other hand,
Madelung's rules do apply to an incredible number of magnetic
systems under a vast range of conditions.

Now that we know that the microscopic magnetic
domains are intimately involved in the production
of magnetic hysteresis, we
immediately come to the first question at the
core of our investigation:
How do the magnetic domains behave on the microscopic level.
Do the domains remember---{\it {viz.}}, return precisely
to---their initial
states, or does just the ensemble average remember?
We show below that, at room temperature, some of the
domains in our samples return to their original configurations and some
do not, but nevertheless the
macroscopic magnetization---set by the ensemble
average---does return to its original value.

In other words, we find that our samples have perfect macroscopic
RPM, but they have imperfect microscopic RPM at room temperature.
In fact, our measured
RPM values for each sample demonstrate a rich, complex behavior
reflecting the fundamental physics of the magnetic domains.  We
quantitatively measured the fraction of the domains that remember
and thereby demonstrated that the disorder has a profound
impact on the microscopic RPM. As we tune the disorder,
our samples develop microscopic RPM that starts from zero
in the low-disorder limit and jumps to a saturated value in the high-disorder limit, but never becomes perfect at room temperature.
Consequently, our experimental system is a finite-temperature realization of the 
``microscopic disorder-induced phase transition between no memory 
and perfect memory" predicted by Sethan, Dahmen, and coworkers \cite{sethna-dahmen}.

The major loop for ``any typical" magnetic system usually has an
additional symmetry---it is symmetric about inversion through the 
origin.
This inversion symmetry immediately raises the second question at the
core of our investigation:  How are the domains at the complementary
points
of the major loop related? Do the magnetic domains at the opposing
points
on the major loop evolve in a similar, but perhaps mirror correlated
fashion?  We call this effect microscopic major loop CPM.
The geometry of complementary-point memory
is illustrated in Fig. 2.

Despite an incredible amount of effort since 1905,
it has proven impossible to develop a simple,
yet adequate, phenomenological model that can be
used to treat all magnetic materials.
We still do not have a phenomenological model for
modern magnetic technology.
In addition, although there has also been tremendous effort
expended and progress achieved,
it has similarly proven impossible to develop a general purpose
micromagnetics model.  We now know, based on recent theoretical work
\cite{sethna-dahmen}, that the detailed
magnetic hysteresis properties of real materials cannot be treated
using standard mean-field methods.  This is because the hysteresis
depends on the interactions between each domain and a limited number
of its neighbors, as well as between each domain and its local disorder.
Consequently, our approach has been to determine to what extent
the nanoscale domain-level physics of our experimental
system obeys Madelung's rules, and then to explore whether we
can better understand the observed behavior using traditional
(overly) simplified Ising models.

\begin{figure}
\includegraphics[width=8.5cm]{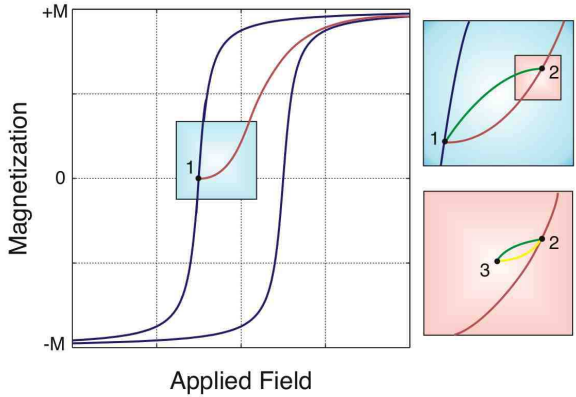}
\caption{The topology of Madelung's rules.
(a) If the magnetization curves from point 1 to saturation are uniquely 
determined by the applied field at
their departure point 1 from
the major loop, then the system exhibits macroscopic major-loop return-point memory.  
If after returning to point 1 on the major hysteresis
loop, the system continues along the major loop, then the system
exhibits macroscopic major-loop memory deletion (wiping out).
(b) If the first-order reversal curve from point 2 back to the major
loop arrives at its original departure point, then the system exhibits first-order macroscopic 
minor-loop return-point memory.
(c) If the second-order reversal curve from point 3 back towards
saturation passes through point 2, then the system exhibits second-order macroscopic 
minor-loop return-point memory.  If thereafter it continues along the original curve 
from 1 to saturation, then the system
exhibits macroscopic minor-loop memory deletion (wiping out).
}
\end{figure}

\begin{figure}
\includegraphics[width=8.5cm]{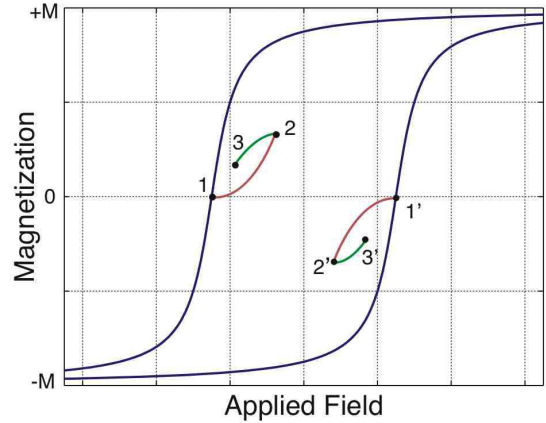}
\caption{The Geometry of complementary-point memory (CPM).
(a) If the magnetization at point $1^{\prime}$ is equal to that at 
point 1,
then the system exhibits macroscopic major-loop CPM.
If the domain configuration at point $1^{\prime}$ is highly
correlated with that at point 1, then the system exhibits
microscopic major-loop CPM.
(b) If the magnetization at point $2^{\prime}$ is equal to 
that at point 2,
then the system exhibits first-order macroscopic minor-loop
CPM.  If the domain configuration at
point $2^{\prime}$ is highly correlated with that at point 2, then the
system exhibits first-order microscopic
minor-loop CPM.
(c) If the magnetization at point $3^{\prime}$ is equal to that at point 3,
then the system exhibits macroscopic second-order minor-loop
CPM.  If the domain configuration at
point $3^{\prime}$ is highly correlated with that at point 3, then the
system exhibits second-order microscopic
minor-loop CPM.
In general, CPM can occur for any order of reversal.
}
\end{figure}

\section{Experimental Aspects}

\begin{figure}
\includegraphics[width = 9cm]{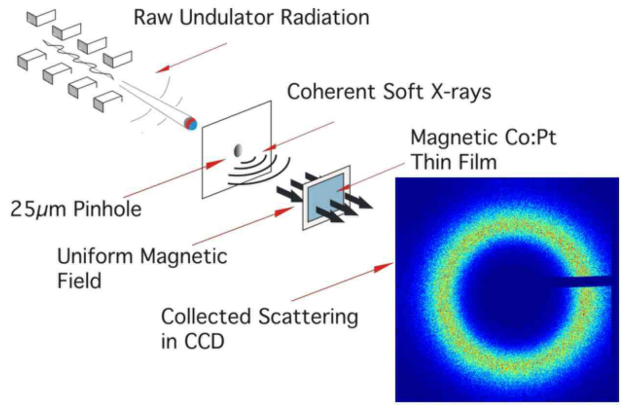}
\caption{(color online)  Schematic diagram of the experimental apparatus.
Soft x-rays from the undulator passed through a pinhole and
were perpendicularly incident on the thin film samples.  The x-rays
were scattered in transmisson and were detected by a soft x-ray
CCD camera.  Not shown in this diagram is the electromagnet used
to apply uniform magnetic fields perpendicular to the sample.
}
\label{fg:experimental_equip}
\end{figure}

To measure the field-history induced changes in the microscopic  
magnetic domain configurations, we developed coherent x-ray speckle  
metrology (CXSM) \cite{prl1, prl2, srn}.
Our CXSM experiments were performed at the Advanced Light Source at  
Lawrence
Berkeley National Laboratory.  A schematic diagram of the experimental  
apparatus is shown in Figure~\ref{fg:experimental_equip}.  We used  
linearly polarized x-rays
from the third and higher harmonics of the beamline 9 undulator.  The  
raw undulator
beam was first reflected from a nickel-coated-bremmstrahlung-safety  
mirror and then
passed through a water-cooled Be window to decrease unwanted light. The  
partially coherent incident beam from the undulator
was passed through a 35-micron-diameter pinhole to select a transversely  
coherent portion.  The sample was located 40 centimeters  
downstream of the coherence-selection
pinhole. This provided transversely coherent illumination of about a 40  
micron
diameter area of the sample.  The transversely coherent x-ray beam was  
incident
perpendicular to the sample surface and was scattered in transmission  
by the
sample.  The resonant magnetic scattering was detected by a soft x-ray  
CCD
camera located 1.1 meters downstream of the sample.  Between the sample
and the CCD camera we used a small blocker to prevent the direct beam  
from damaging the CCD.

The photon energy was set to the cobalt L3 resonance at $ 778$ eV.   
These
photons resonantly excited virtual 2p to 3d transitions in the cobalt atoms and  
thereby
provided our magnetic sensitivity.  The intensity of the raw undulator  
beam was
$2 \times 10^{14}$ photons/sec, the intensity of the coherent beam was
$2 \times 10^{12}$ photons/sec, and the intensity of the scattered beam
was $2 \times 10^7$ photons/sec.  We typically measured each speckle
pattern for 10 to 100 seconds, so the total number of photons in each  
CCD
image was $10^{8}$ to $10^{9}$.

The applied magnetic field was provided by an in-vacuum water-cooled
electromagnet allowing {\it in situ} adjustment of the magnetic field during the experiment.  The return path for the electromagnet consists of an external soft Fe yoke that feeds field to vanadium permandur pole pieces that are integral to the vacuum chamber.  The pressure inside the
chamber during our experiments was typically $10^{-8}$ Torr.
The in-vacuum electromagnet provided magnetic fields up to $11$ kOe.

\subsection{Sample Fabrication and Structural Characterization}

\begin{figure}
\includegraphics[width = 8.75cm]{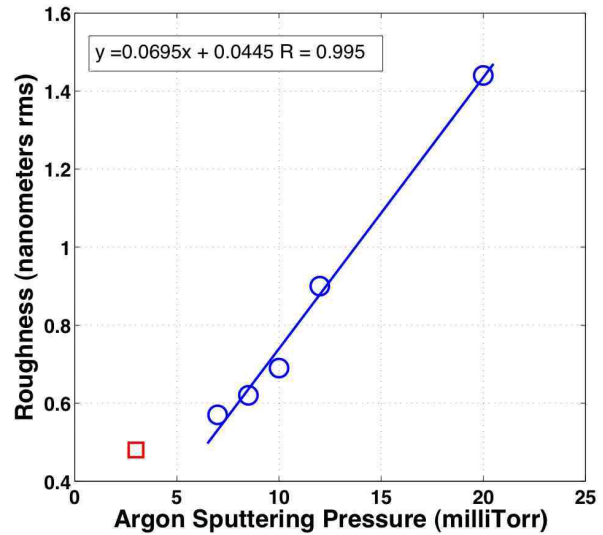}
\caption{(color online) The measured {\it {rms}} roughness 
from AFM and x-ray reflectivity  
measurements plotted versus the Argon sputtering pressures.  The interfacial roughness 
increases as the sample growth pressure increases.  Below 8.5mTorr, the roughness increases slowly as indicated by the left (green) fit line.  Above 8.5mTorrthe roughness increases much more rapidly as indicated by the included by the right (red) fit line.  This behavior is very similar to that observed by Ref. \cite{eric-prb} in sputtered Nb/Si multilayers.
}
\label{fg:pressure_roughness}
\end{figure}

\begin{figure*}
\includegraphics[width=18cm]{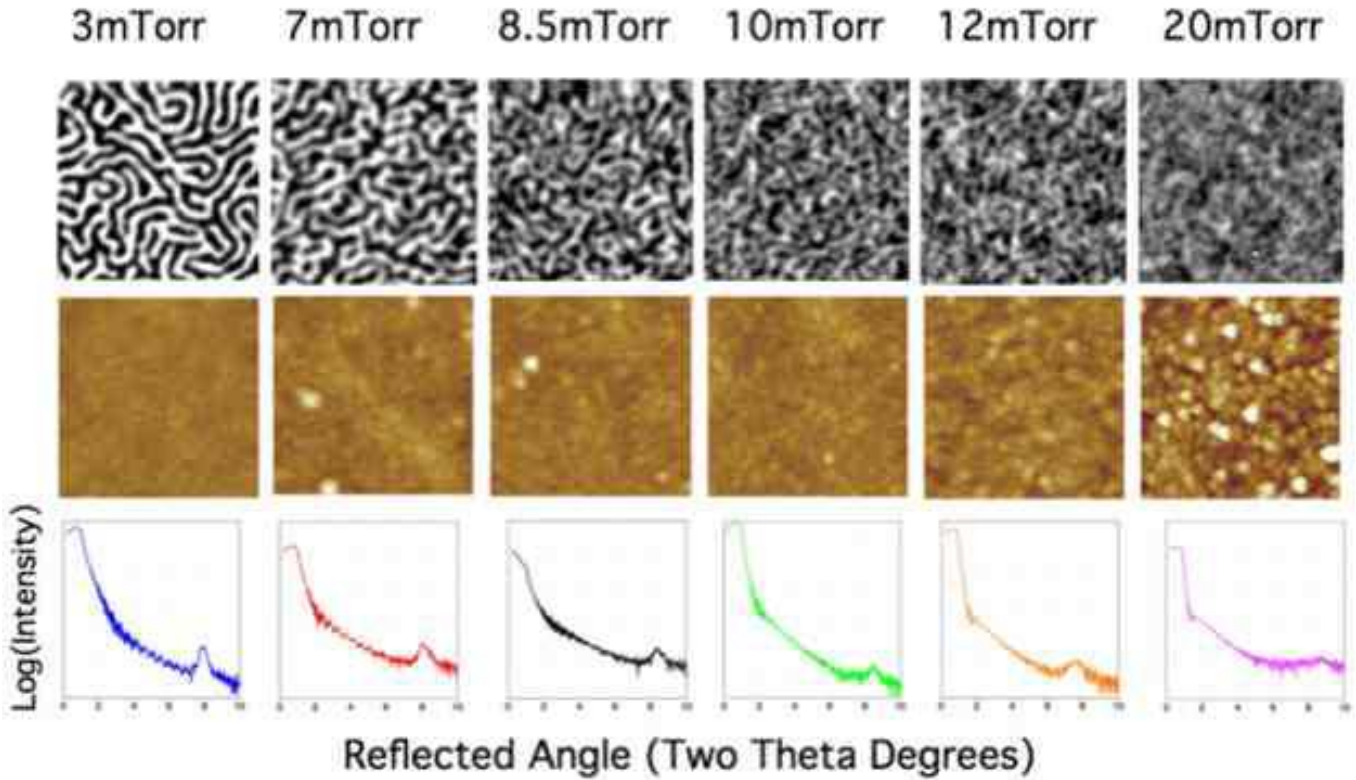}
\caption{(color online) The measured MFM images, AFM images, and x-ray reflectivity 
curves for the six samples.  The MFM images evolve from clear labyrinthine 
patterns for the low {\it {rms}} roughness samples to visually highly disordered patterns 
for the high {\it {rms}} roughness samples. However, the persistence 
of the annular
shape of the speckle patterns---even for the highest roughness 
samples---reveals an
underlying labyrinthine order. 
The MFM images show 3 micron by 3 micron areas. 
The AFM images show that the top surface of the samples becomes more  
rougher at higher pressures. 
The AFM images show 1 micron by 1 micron areas.
Both the x-ray reflectivity curves and the 
AFM images were used to determine the {\it {rms}} roughness for each sample.
}
\label{fg:6_sisters}
\end{figure*}

Our thin-film samples were grown by magnetron  
sputtering in the San Jose Hitachi Global Storage  
Technology Laboratory
on smooth, low-stress, 160-nm-thick silicon nitride membranes.  The samples had
20-nm-thick Pt buffer layers, and 2.3-nm-thick Pt caps to  
prevent oxidation.  Between
the buffer layer and the cap, the samples had 50 repeating units of a  
0.4-nm-thick Co layer and
a 0.7-nm-thick Pt layer.  While the six samples had identical  
multilayer structure they were
grown at different argon sputtering pressures to tune the disorder in  
the samples.
During growth, we adjusted the deposition times to keep the Co and Pt  
layer thickness constant
over the entire series.  For low argon pressures, the  
sputtered metal atoms arrive at the
growth substrate with considerable kinetic energy which locally heats  
and anneals the growing
film.
This leads to smooth Co/Pt interfaces produced at a low sputtering  
pressure.
For higher argon sputtering pressures, the sputtered atoms arrive at  
the growth substrate with
minimal kinetic energy thereby resulting in rougher Co/Pt
interfaces.  The resulting roughness is cumulative through the samples \cite{eric-prb}.
The magnetocrystaline anisotropy at the Co/Pt interface forces
the magnetization to align perpendicularly to the surface of the film.
Our samples were grown at six different sputtering pressures:
3, 7, 8.5, 10, 12, and 20 mTorr.
Due to the important and interesting magnetic properties, these samples and others very similar in form and structure have been studied in different experiments\cite{olav_physicab, davies}.
%\footnote{To avoid confusion between pressure and magnetic field units, we always use Oe for magnetic field units and never use Tesla.}.
%In this paper, we will identify each sample based on its growth
%pressure in mTorr (milli-Torr).  To avoid confusion, Oe and kOe are used for magnetic field units while %the unit of Tesla is never used.

The {\it rms} roughness for the samples was measured in 
the Almaden Hitachi Global Storage
Technology Laboratory using two different methods.  First, we measured  
the roughness by
scanning the sample surface with an atomic force microscope (AFM)
and calculating the {\it rms} roughness from the AFM images.  Since our
samples have conformal roughness, the {\it rms} roughness of the surface
is a reasonable measure of the internal {\it {rms}} roughness.  However,
to directly probe the internal {\it rms} roughness, we also did 
the x-ray reflectivity measurements shown in Fig. \ref{fg:6_sisters}.  The reflectivity data was fit using a Debye-Waller factor to determine the roughness.  Instead of the system possessing thermal fluctuations, the displacements from the average height are randomly distributed and fixed.
The {\it rms} roughness values from the x-ray measurements
agreed with those from the AFM measurements, confirming
the conformal roughness of our samples.
The {\it rms} roughness values are shown in 
Fig. \ref{fg:pressure_roughness} and are listed in Table 1.
We found that the {\it rms} roughness for the 3 mTorr sample
is about 0.48 nm and that it increases to 1.44 nm for the
20 mTorr sample.

\subsection{Magnetic Characterization
}

\begin{table}
\caption{\label{tab:table1} The Measured Magnetic Characteristics of Our Six Samples }
\begin{ruledtabular}
\begin{tabular}{|l|cc|ccr|}
Sample\footnote{Our samples are labeled by their growth pressure in mTorr}
& $\sigma_{{\it {rms}}}$ \footnote{The measured {\it {rms}} interfacial roughness in nm}
& $M_{s}$ \footnote{The measured saturation magnetization of Co in $emu/cm^{3}$}
& $H_{n}$ \footnote{The nucleation field measured from positive saturation}
& $H_{c}$ \footnote{The measured coercive field in kOe}
& $H_{s}$ \footnote{The measured saturation field in kOe}\\
\hline
3 mTorr   & 0.48  & 1360  & 1.58   & 0.16  & 3.7\\
7 mTorr   & 0.57  & 1392  & 0.64   & 0.68  & 5.0\\
8.5 mTorr & 0.62  & 1136  & 1.68   & 1.42  & 5.5\\
10 mTorr  & 0.69  & 1069  & 1.45   & 1.87  & 6.5\\
12 mTorr  & 0.90  & 1101  & 1.23   & 2.74  & 9.5\\
20 mTorr  & 1.44  & 918   & -1.81  & 5.89  & 14.2\\

\end{tabular}
\end{ruledtabular}

\end{table}

\begin{figure}
\includegraphics[width=8.5cm]{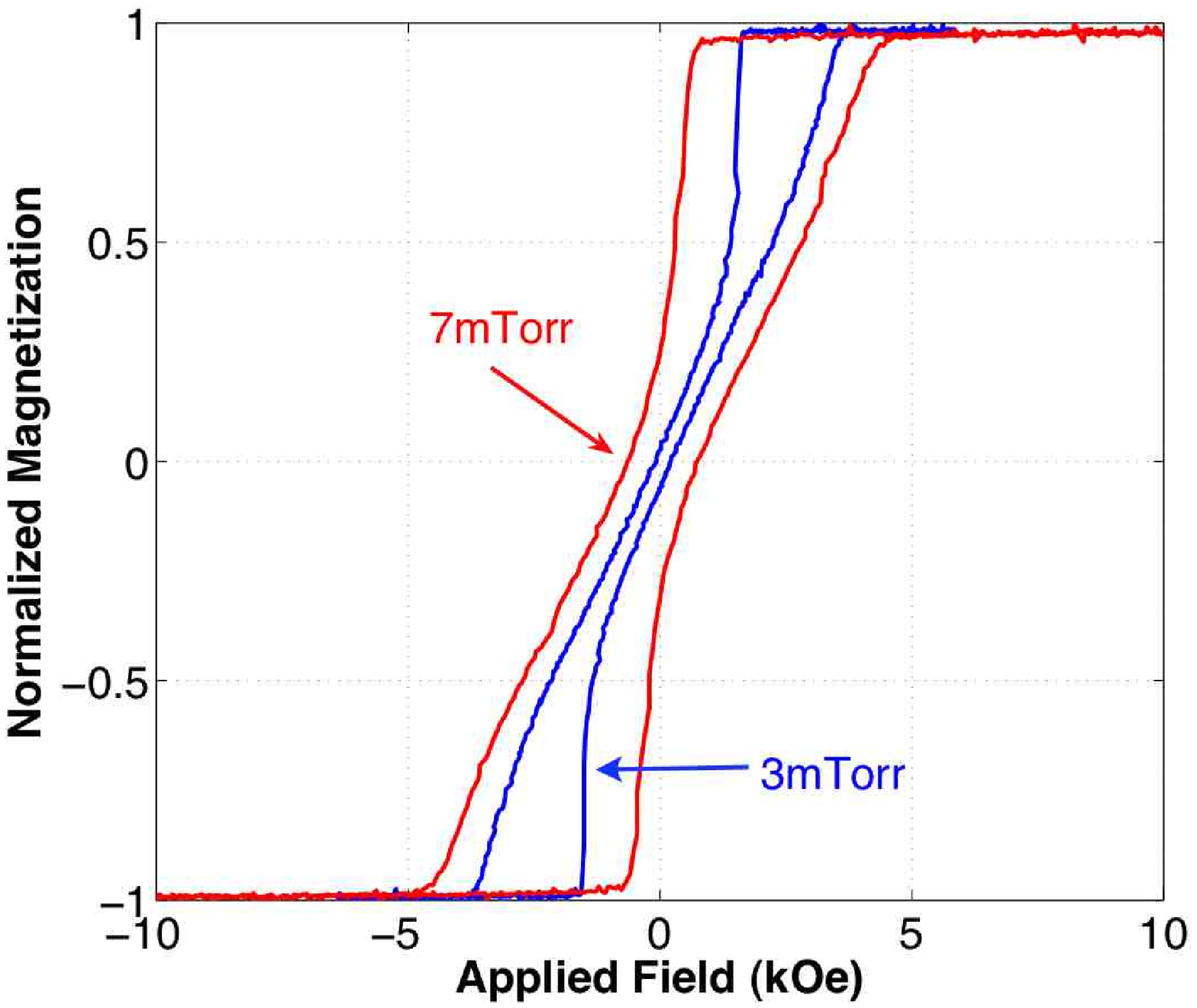}
\includegraphics[width=8.5cm]{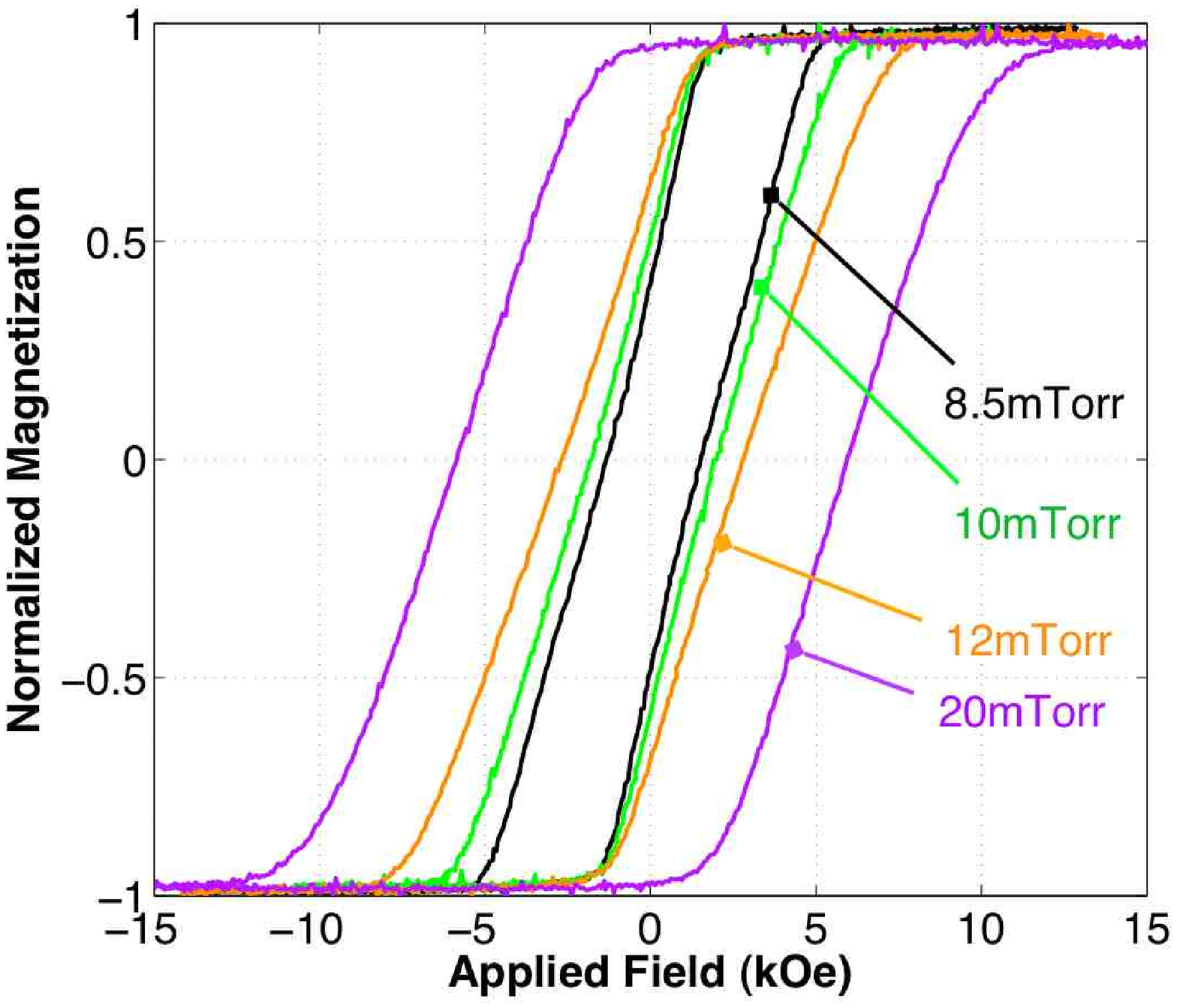}
\caption{(color online)  The measured magnetic hysteresis
loops for our samples.  
Note that the shape of the major hysteresis loops change abruptly
above the ``critical roughness" value---which occurs between the 7
and 8.5 mTorr samples---and that the areas inside the major loop
increase as the disorder increases past the ``critical roughness" value.
The two low {\it rms} roughness samples possess exhibit ``classic
Kooy-Enz behavior" characterized by a sharp nucleation region
and low remnant magnetization, 
whereas the high {\it rms} roughness samples exhibit an almost 
constant slope.
}
\label{fg:majorloop}
\end{figure}

\begin{figure}
\includegraphics[width=8.5cm]{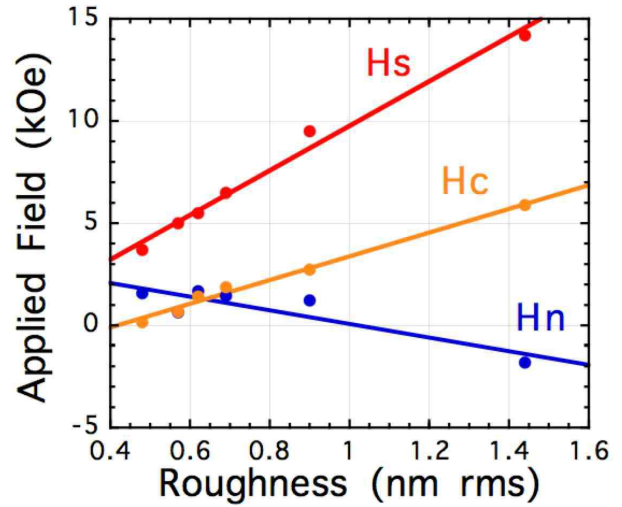}
\caption{(color online) The measured magnetic characteristics for our samples 
plotted versus their measured {\it rms} roughness.  
The coercive, nucleation, and saturation fields are denoted by
$H_c$, $H_n$, and $H_s$, respectively.
Note the apparently
linear dependence of these properties on the {\it rms} roughness.
}
\label{fg:magnetic_character}
\end{figure}

We measured the major hysteresis loops
for all of our samples using both Kerr magnetometry at the San Jose  
Hitachi Global Storage Technology Laboratory
and alternating gradient magnetometry (AGM) at the University of 
California-Davis. The
measured major loops shown in Fig. \ref{fg:majorloop} 
exhibit clear changes that are related
to the increasing roughness.  The two low disorder films
(3 mTorr and 7 mTorr) exhibited ``classic Kooy-Enz \cite{kooy-enz}
soft loops" with
low remanence and abrupt nucleation transitions.
Between 7 mTorr and 8.5 mTorr, there is
an abrupt transition to loops which do not show
a clear nucleation region.  Between 8.5 mTorr and 20 mTorr,
the ascending and descending slopes of the loop remain
the approximately the same, but the loops gradually widen until
the full magnetic moment is left at remanence. The values of the nucleation, coercive and saturation fields each exhibit a roughly linear dependence upon the sample roughness. This behavior is shown in Fig. \ref{fg:magnetic_character}. 
In addition, we also found via magnetometry that
all of our films exhibit perfect macroscopic
{\it{major loop and minor loop}} RPM and CPM.

The AGM magnetometer was used to measure the  
saturation magnetization of the samples.
The measured saturation magnetizations are reported in
Table 1; they should be compared against the value of
$M_{s} = 1400 {\rm emu/cm}^3$ for pure cobalt.
It is interesting to note that with increasing interfacial roughness,
that the coercivity and saturation field increase and the nucleation
field decreases linearly with the roughness;
the measured saturation magnetizations also 
decrease as the disorder increases.

\subsection{Coherent X-Ray Speckle Metrology}

To measure the field-history induced changes in the 
correlations between the microscopic
magnetic domain configurations, we developed coherent x-ray speckle
metrology (CXSM).
The magnetic sensitivity of CXSM is provided by virtual 2p to 3d
resonant magnetic scattering.
We produce a transversely coherent beam by passing the 
partially coherent beam from the undulator
through a 35-micron-diameter circular pinhole to select a highly transversely
coherent portion. The beam selected by the spatial filter is largely coherent over the entire illuminated area.
Due to this large uniform transverse coherence, the resonant magnetic scattering
produces the speckle patterns that we use to track the
field-history-induced evolution of the magnetic domains.  We explain
our analysis methodology for the magnetic speckle patterns in the next
section.

What information does x-ray speckle metrology provide about the
magnetic domains, and why don't we simply study the magnetic domains
in real space?  This is the venerable old question about diffraction
versus microscopy.  The conventional answer is that they are
complementary: use ``conventional beam diffraction" to obtain
information about the ensemble average, and use microscopy to
obtain information about the individual defects. There are two
limiting cases of conventional diffraction studies.
When the diffraction
pattern consists of Bragg peaks, then the information that 
conventional diffraction provides is the ensemble average of the
long-range order.
When the diffraction
pattern consists of diffuse scattering, then the information that 
conventional diffraction provides is the ensemble average of the
short-range order.
In our labyrinthine systems, there is no long-range magnetic order, 
and consequently the diffraction is diffuse.  With an unfiltered beam we observe only a diffuse annulus which contains
information about the strength (amplitude) of the magnetic domains,
the mean spacing of the magnetic domains, and the
correlation length of the magnetic domains.  We have performed such studies already and the results are in preparation for publication.

Fully coherent diffraction
changes that paradigm because the precise configuration of the
speckles provide detailed information about the defects, or in
our case about the configuration of the magnetic labyrinths.
In the
Bragg case, the information is about the defects in the crystalline
order.  In the diffuse case, the information is about the defects
in the short-range order.
In fact,
in two- or three-dimensions, if the speckle pattern is sampled with
sufficient wavevector resolution, then all of the real-space 
information is contained in the speckle pattern and
can be recovered using ``oversampling speckle reconstruction" \cite{miao}.
However, no successful oversampling speckle reconstructions
have yet been reported for magnetic domains, though holography methods have recently been demonstrated in similar systems\cite{luening_speckle}.
Consequently, our objective was not to extract the complete real-space 
information,
but instead to directly determine the changes between the correlations
of the magnetic 
domain configurations prepared via different applied-field 
histories---specifically without requiring the oversampling speckle 
reconstruction of our speckle image magnetic fingerprints.

So what information does our magnetic speckle metrology provide?
We illuminate a 40 micron diameter circle on the sample so our
ensemble-average is over that region.  Consequently, each speckle
in our speckle pattern consists of an Airy pattern with a characteristic
size in reciprocal space of $2 \pi / 40$ inverse microns.  The
second important length scale in our problem is set by the width
of the magnetic domains in the labyrinth state.  This width is
200 nm, and consequently the corresponding characteristic
size in reciprocal space is $2 \pi / 200$ inverse nm;
this sets the mean-radius of our annular speckle patterns.  In
principle, our speckle patterns can provide spatial information down
to $\lambda/2 = 0.8$ nm, but in practice---due to the strong disorder in
our labyrinths---our speckle patterns really only contain information
set by the region where the diffuse scattering is measurable,
namely between the inside and outside radii of the annulus.
For our samples this was from about
110 to 260 nm in real space.

As argued above,
all of the physical information that can be obtained using our
incident wavelength is contained within the limited range that
contains measurable scattering intensity.
Our incident wavelength is fixed by the magnetic resonant scattering 
condition for cobalt so $\lambda \simeq 1.6$ nm.
For this wavelength, diffraction provides information
ranging from 0.8 nm for backscattering with $2\theta = 180$
degrees up to 40 microns set by the illumination area.
At our usual sample-to-camera separation,
the pixel size of our camera translates into a real-space
resolution of 13 microns and the total coverage of the camera
translates into a real-space resolution of 270 nm.
The angular size of our beamstop translates into 70 nm.
Since $70 < 110$ nm, $260 < 270$ nm,
and $27 < 40$ microns,
our camera and our beamstop do not limit the spatial 
scales that we can access.   
Instead, the limits are set only by the disorder levels in our samples.

The intensity $I(q_r, q_{\theta})$ of each speckle located 
at position $(q_r, q_{\theta})$ is
proportional to the square-modulus $  \mid a_q \mid^2 $
of the scattering amplitude $a_q$
of the corresponding Fourier component of the magnetic
density $\rho_{\rm mag}(q_r, q_{\theta})$.  So by taking the
square root
of the intensity of our speckle pattern, we can first calculate and then visualize
the result as a map of the magnetic density amplitudes for
all of the most important Fourier components.  Each component
located at ${\bf q} = (q_r, q_{\theta})$
tells us the amplitude of an infinite-spatial-extent complex-valued exponential
density component $\exp(i{\bf q} \cdot {\bf r})$ multiplied by 
our illumination function which is roughly equal to one inside
the illumination circle and zero outside.  Imagine a large
number of these complex-valued oscillating exponentials, each one oriented
along the direction $\theta$ with amplitude 
$\sqrt{I}$
and with wavevector $q_{r}$.

So, how many of these Fourier components do we measure?  The
area of our observed annulus in reciprocal space is given by

$$A_q = \pi {{q^{2}_{\rm {max}} }} - \pi {{q^{2}_{\rm {min}}}} $$

\noindent
and the area of each one of our speckles in reciprocal space is given by

$$A_{\rm{speckle}} = \pi {\delta q_{\rm {speckle}}^{2}} $$

\noindent
so the number of speckles inside the annulus is given by

$$A_{\rm{annulus}} / A_{\rm{speckle}} \; \simeq \; 30,000. $$

\noindent
In other words, we directly measure this many Fourier components
of the magnetization density.  Because the speckle intensity
outside the annulus is negligible, the corresponding
Fourier components outside the annulus are also negligible.
So we directly obtain information about all of the 
non-negligible Fourier components of the magnetic density
that produce the magnetic scattering within the speckled annular
region that we measure.

On the other hand, modern
computer control and computer image analysis should enable 
modern magnetic x-ray microscopy
to obtain ensemble-averaged information about the magnetic
domains.  This is certainly worth pursuing, and we are just
beginning such studies.

Speckle contrast, the normalized standard deviation of the intensity, is generally used as a measure of the quality of the produced speckle patterns.  As the diffuse scattering envelope is azimuthally symmetric about $q_r = 0$, it is correct to define the speckle contrast $\sigma^2_{con}(q_r)$ as

$$\sigma_{con}^2(q_r) = \frac{1}{\langle I \rangle}{\sqrt{\sum_k^N \frac{(I_k - \langle I \rangle)^2}{N-1}}}$$

\noindent
for small steps of $q_r$ where the sum is carried out over all $N$ is the number of data points included in each step.
 Using this calculation, the contrast present in our speckle patterns typically ranges from $0.6$ to $0.4$, with a small dip in values over the peak scattering.  This interesting variation of the speckle contrast as it depends upon $q_r$ is quite reminiscent of the speckle contrast studies by Retsch and McNulty \cite{mcnulty_prl} across absorption edges and could provide useful information if properly understood.

\section{Data and Data Analysis
}

\begin{figure}
\includegraphics[width=8cm]{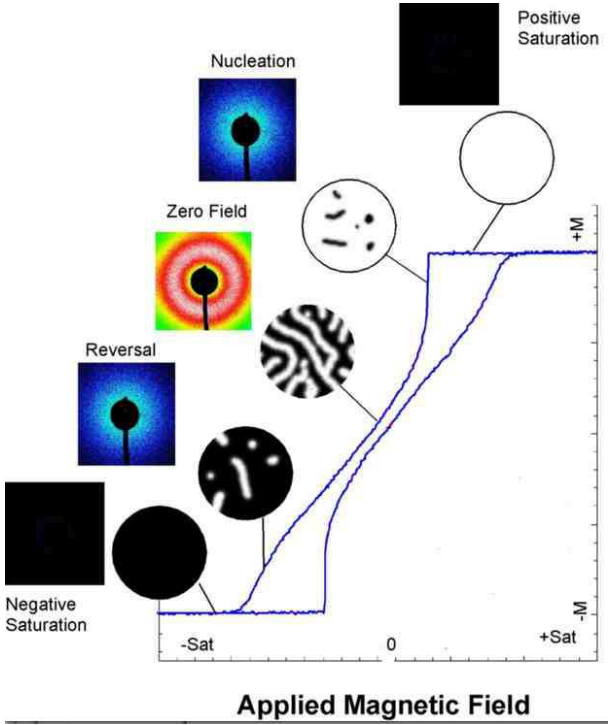}
\caption{(color online) The measured major hysteresis loop for the 3 mTorr sample.   
The measured magnetic speckle patterns collected at different  
field values along the major loop 
are shown inside the square insets and 
an artist's rendition of the corresponding 
magnetic domain configurations are shown inside the circular insets.
}
\label{fg:speckle_hysteresis}
\end{figure}

\begin{figure}
\includegraphics[width=8cm]{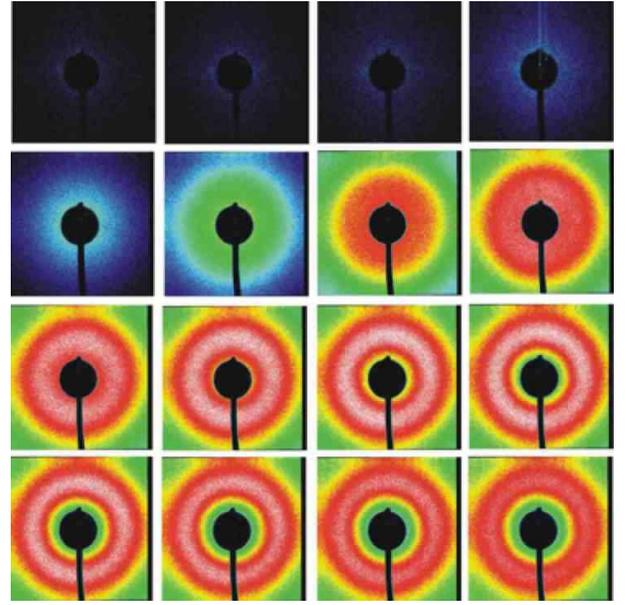}
\caption{(color online) The measured CCD magnetic speckle images 
for about the first half of the ascending major loop for the low disorder 3 mTorr sample.
The dark region in the center is a blocker inserted to eliminate the
direct beam from the image.
These magnetic speckle patterns were collected at fixed values of the  
applied field as the field was monotonically increased in steps from negative saturation.  
The speckle images associated with the applied magnetic field values 
are shown---from left to right and from top to bottom---for the following
applied field values -3.0, -2.5, -2.0, -1.75, -1.5, -1.25, 
-1, -0.75, -0.5, -0.25, 0, 0.25, 0.5, 0.75, 1, and 1.25 kOe.  
}
\label{fg:donuts}
\end{figure}

The typical evolution of the speckle patterns for one-half cycle around
the major hysteresis loop for the 3 sample is schematically
illustrated in Fig. \ref{fg:speckle_hysteresis}.  
and the corresponding measured speckle patterns are shown in Fig. \ref{fg:donuts}.
Starting at positive saturation there is no magnetic contrast---all
the magnetic domains are aligned with the field---consequently
there is no magnetic scattering.  As we descend from positive saturation
the magnetic domains nucleate and produce a magnetic speckle pattern
that is shaped like a cookie (disk).  When we reach zero applied field,
the magnetic domains have grown so much that they fill the entire
sample; in this limit they must interact, and their interaction produces
the donut (annular) shaped speckle pattern.  When we reach the reversal
region, the domain density is again low, and so the associated
speckle pattern is again cookie shaped.

\subsection{Correlation Coefficients}

To quantitatively compare the magnetic domain configurations versus the
applied magnetic field history, we calculate the normalized correlation  
coefficients between pairs of our measured images acquired for different
applied field histories.  To date, our work has been primarily based on 
the normalized cross-correlation of these magnetic speckle patterns---our
magnetic speckle fingerprints.  However, this comparison can be done in
reciprocal space---as we have primarily been doing up until now---or in
real space as we are just beginning to do.

\begin{figure}
\includegraphics[width=4.2cm]{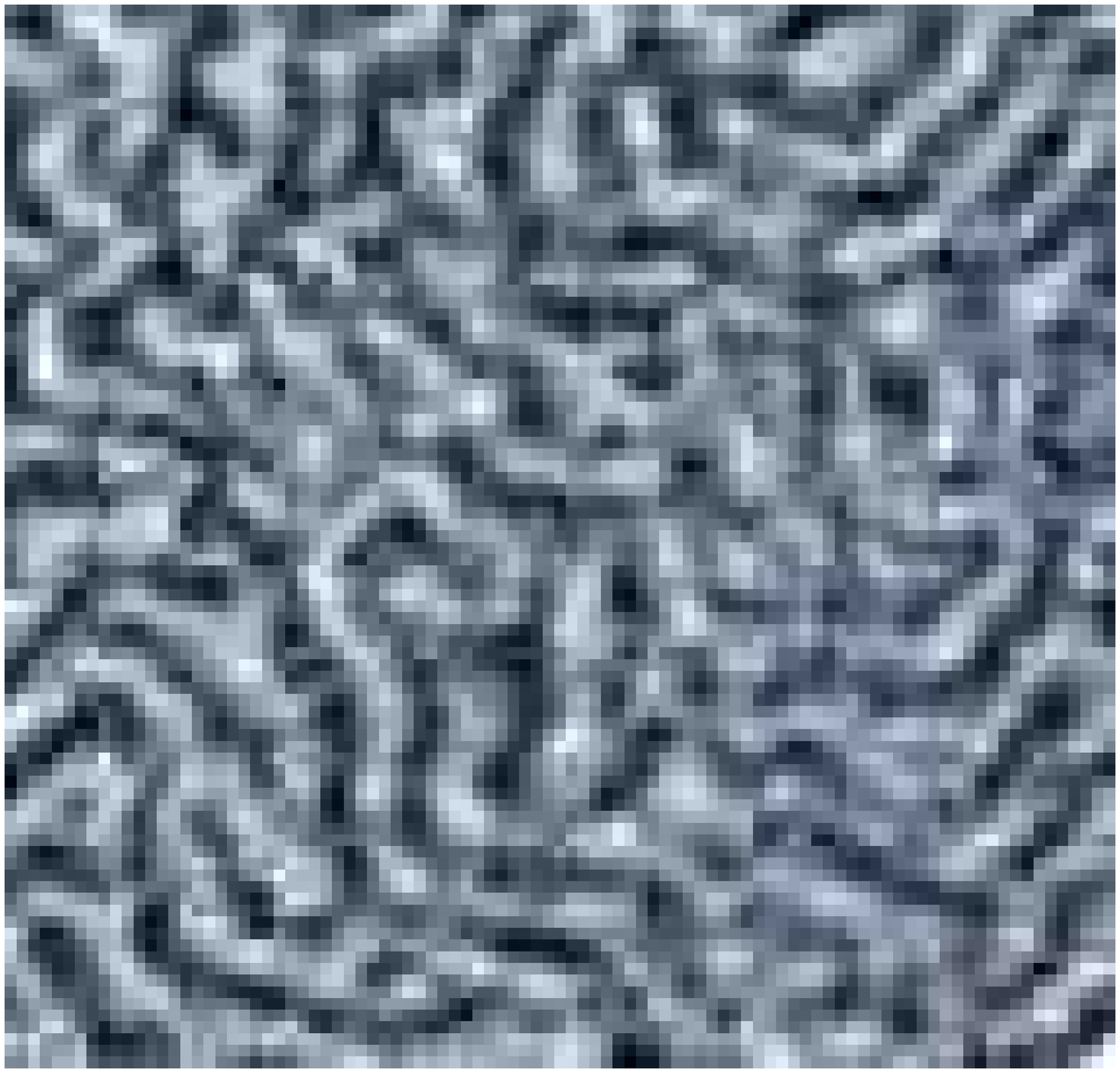}
\includegraphics[width=4.2cm]{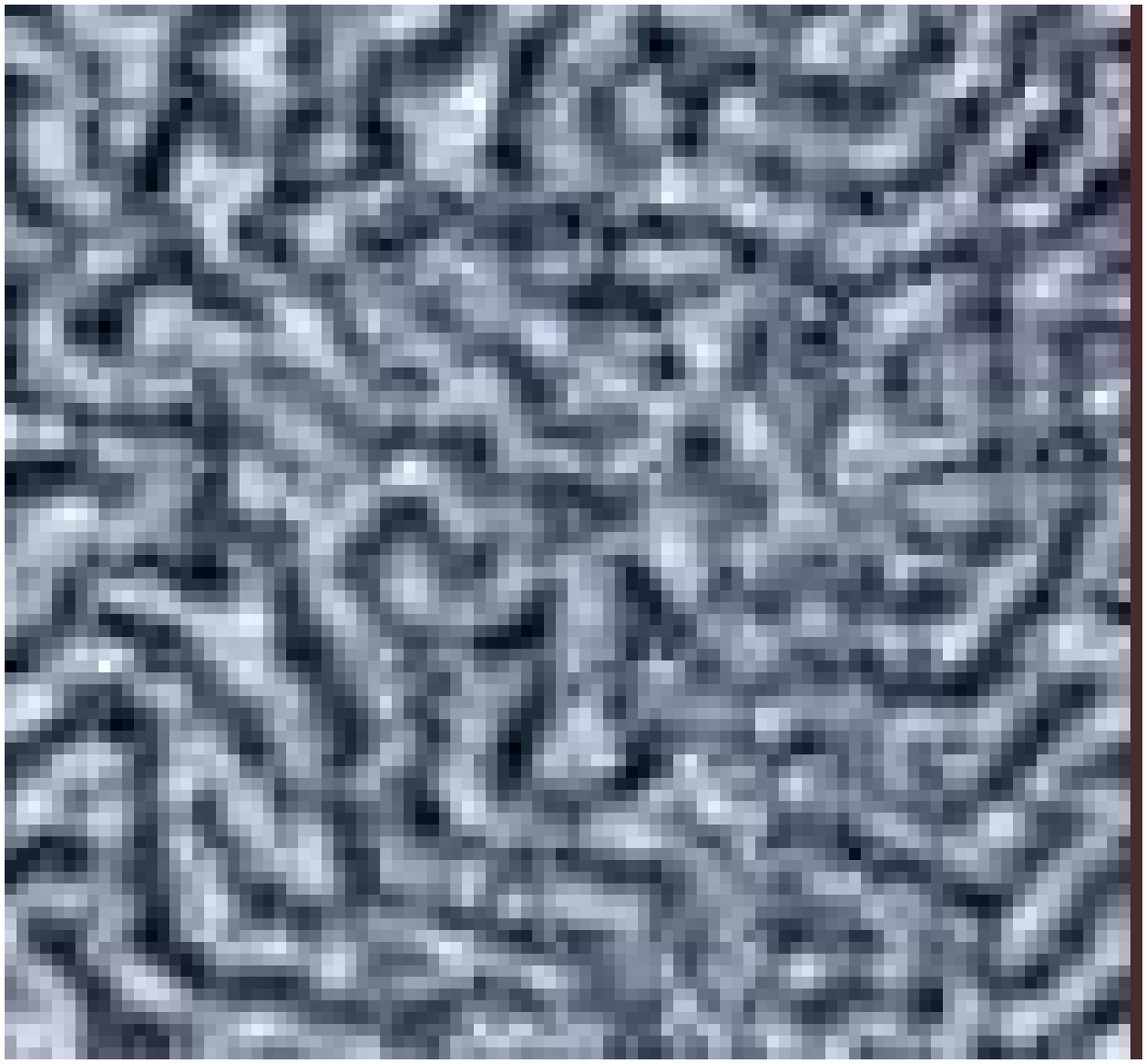}
\caption{(color online) Two representative magnetic x-ray microscopy images 
for our 8.5 mTorr sample 
at $H= -0.50$~kOe (left) and $H=-1.00$~kOe (right).
These images show $2.2 \times 2.2 \mu m$ areas.
}
\label{fg:real_space_85}
\end{figure}

Some of our initial experimental work in real space is illustrated   
Figure \ref{fg:real_space_85} which shows the magnetic domains in
our 8.5 mTorr sample measured using x-ray magnetic 
microscopy \cite{peter RSI}.  These images were recorded at the Co L3 edge 
using XM-1 at the ALS and were taken on the descending major loop;
the left panel shows the domains at $H= -0.50$~kOe and the right panel shows the 
domains at $H=-1.00$~kOe.  The correlation coefficients obtained 
from our real-space normalized cross-correlation analysis of the domain 
patterns agrees with our correlation coefficients obtained via our
standard reciprocal-space normalized cross-correlation analysis of the 
corresponding speckle patterns.
Therefore we believe that our real space and reciprocal space methods
will prove to be complementary.

\begin{figure}
\includegraphics[width=4.2cm]{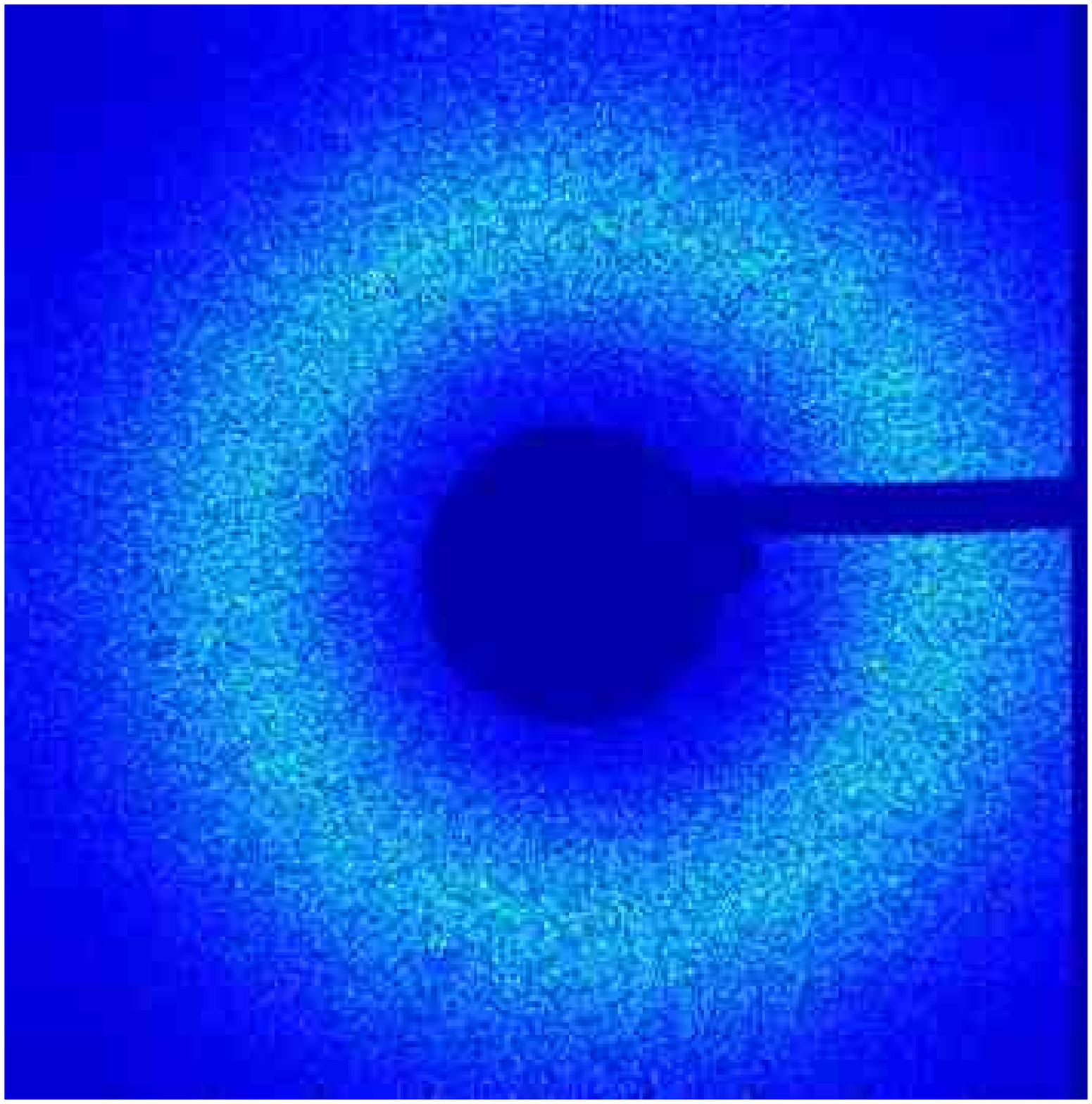}
\includegraphics[width=4.2cm]{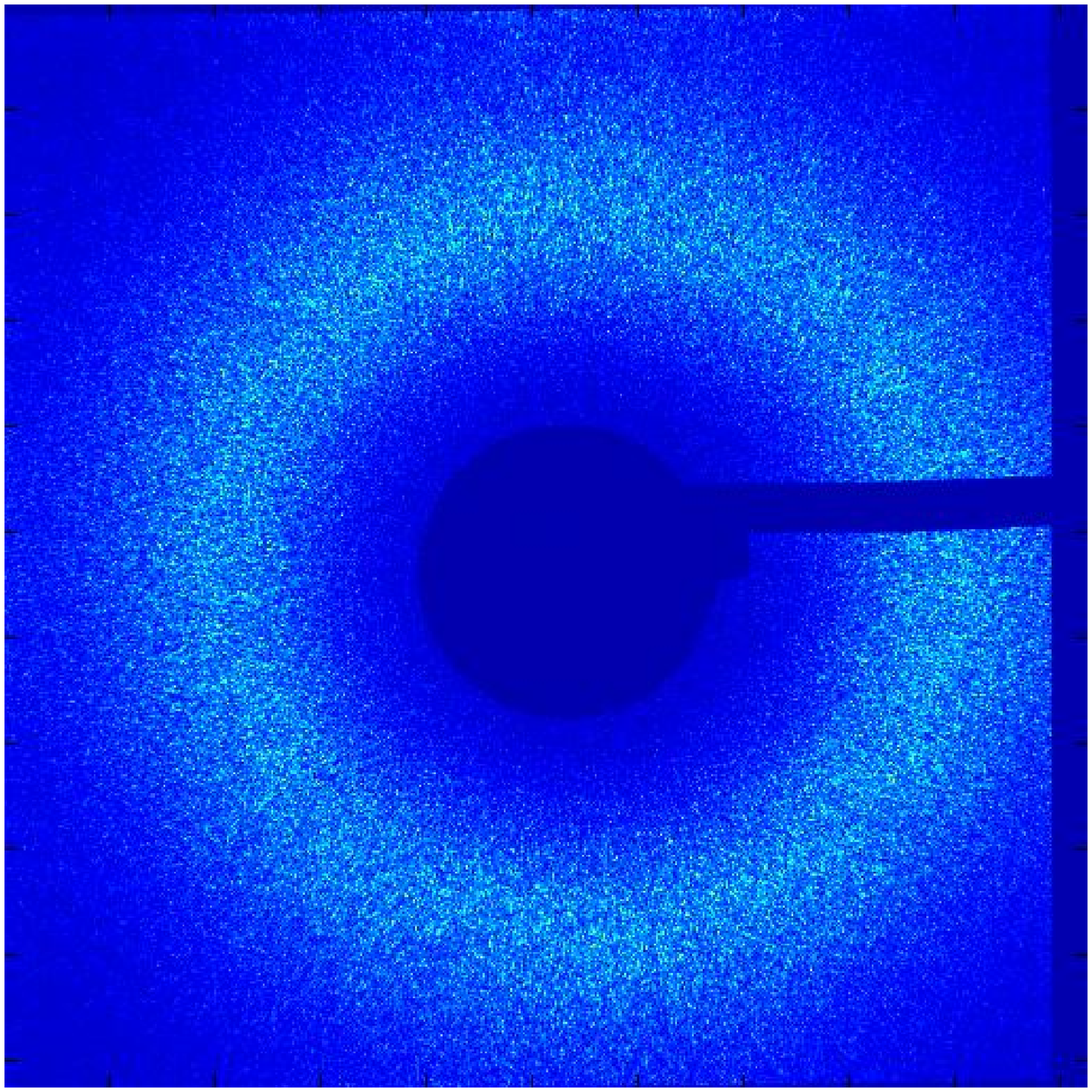}
\caption{(color online) Two magnetic speckle patterns for our 8.5 mTorr sample 
shown for $H=-0.50$ (left) and $H=-1.00$~kOe (right).
}
\label{fg:q_space_85}
\end{figure}

\begin{figure}
\includegraphics[width=4.2cm]{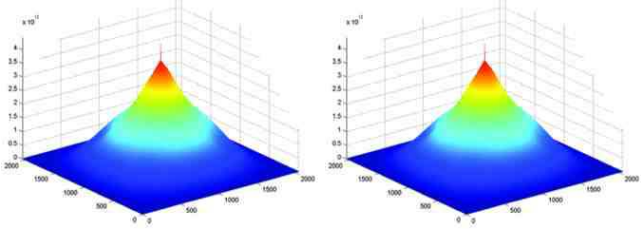}
\includegraphics[width=4.2cm]{cross_2.ps}
\caption{(color online) The two auto-correlation functions for the two magnetic speckle 
patterns shown in Fig. \ref{fg:q_space_85}.
The coherent speckle information is contained in 
the sharp peak that rides on the top of the large 
diffuse scattering signal.
The autocorrelation functions are shown over the 
full 2049 by 2049 pixel area.  The vertical scale 
extends from $0$ to $4 \times 10^{12}$.
}
\label{fg:auto-corr}
\end{figure}

\begin{figure}
\includegraphics[width = 8.5cm]{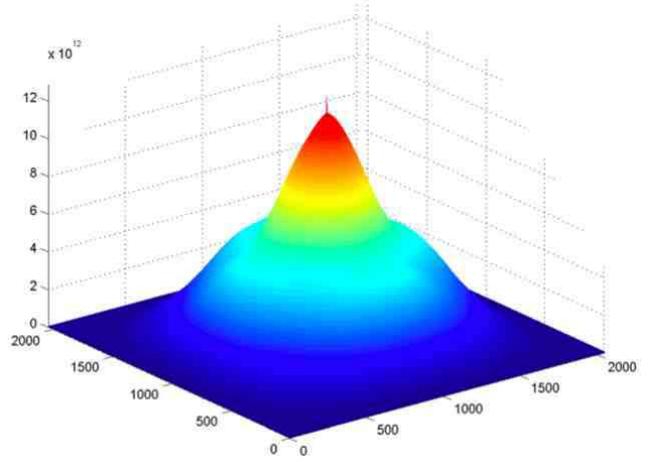}
\caption{(color online) The cross-correlation function between the two speckle patterns shown in Fig. \ref{fg:q_space_85}.  The coherent speckle information is contained in the sharp peak that rides on the top of the large diffuse scattering signal.
The cross-correlation function is shown over the 
full 2049 by 2049 pixel area.  The vertical scale 
extends from $0$ to $4 \times 10^{12}$.
}
\label{fg:corr_toon}
\end{figure}

Our normalized cross-correlation analysis procedure in reciprocal space
is illustrated in figures \ref{fg:q_space_85}-\ref{fg:corr_toon}.
Figure \ref{fg:q_space_85}
shows the speckle fingerprints measured in reciprocal space;
again the left panel  
shows the fingerprint at $H=-0.5$ and the right panel shows the  
fingerprint at $H=-1.0$~kOe.  Figure \ref{fg:auto-corr} shows the calculated  
autocorrelation functions for these two speckle fingerprints.  Note that  
both of these consist of a broad smooth ``mountain" with a sharp ``tree"  
on top of it.  

The mountain corresponds to the diffuse scattering envelope from the short range magnetic ordering 
and the tree corresponds to the coherent scattering from the entire illuminated area. Figure \ref{fg:corr_toon} 
shows the  
calculated cross-correlation function for the two speckle fingerprints
shown in Fig. \ref{fg:auto-corr}.  
Again there is an ``diffuse mountain" with a ``coherent tree" 
on top of it.  

We want to use the coherent components of these auto- and  
cross-correlation functions to
compute the normalized correlation coefficient.  We
extract the volume of each tree and then we calculate the ratio

$$ \rho(a,b)=   \frac {volume(a \otimes b)} { \big[ {volume(a\otimes a)}  \; \; {volume(b \otimes b)} \big] ^{1/2}}. $$

The resulting normalized correlation coefficient $\rho(a, b)$ 
measures
the normalized degree-of-correlation between any two speckle patterns.
We use it to quantify the degree-of-correlation between pairs of speckle
patterns which in turn is our measure of the degree-of-correlation  
between the corresponding magnetic domain patterns.  
When $\rho(a,b)=1$ the
two magnetic domain patterns are identical, and when $\rho(a, b)=0$ the
two magnetic domain patterns are completely different.  

In general, the
value of $\rho$ specifies the degree-of-correlation between
the two speckle patterns which in turn are proportional to
the Fourier coefficients of the magnetization density for
the two magnetic domain configurations.  Because our correlations
are based on the intensity, we are unable to determine the sign of the  
correlations---we cannot distinguish correlation from anti-correlation.  This is essentially Babinet's Principle.

\begin{figure}
\includegraphics[width = 10cm]{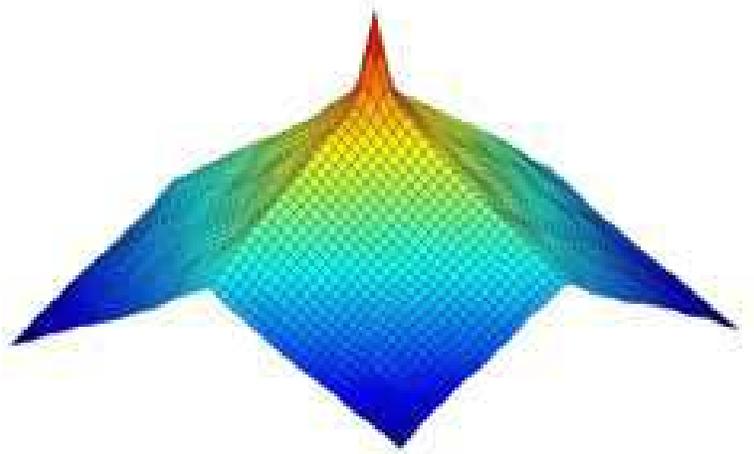}
\caption{(color online) The cross-correlation function 
calculated by comparing small regions in the respective speckle images. 
The coherent speckle signal (the spike shown in yellow) is clearly visible 
and is located at the peak of the large diffuse background 
signal (the roof-shaped structure shown in blue).
}
\label{fg:subtraction_toon}
\end{figure}

\subsection{Data Analysis}

We now sketch the details of our correlation coefficient calculations.
To determine the correlation between two speckle patterns, we
generalized the standard correlation coefficient for two random
variables $a$ and $b$, which is given by

$$\rho(a, b) = {\rm Cov}(a,b) \times \big\{ {\rm Var}(a) {\rm Var}(b)
\big\}^{-1/2} ,$$

\noindent
to the equivalent expression in terms of the
background-subtracted cross-correlation
function (ccf) and the
background-subtracted autocorrelation functions (acf) of
the speckle pattern intensities 
$I_a(q_{x},q_{y})$ and $I_b(q_{x},q_{y})$

$$ \rho(a, b) = \frac{ \sum {\rm ccf}(a,b) }
{ \big[
{ \sum  \;{\rm acf}(a) \; \; \sum 
\; {\rm acf}(b)
} \big]^{1/2} }$$

\noindent
where the sums run only over the trees so that the
background-subtracted auto- and cross-correlation 
functions contain only coherent scattering contributions.

We used the standard fast Fourier transform (FFT) to calculate 
the auto- and cross-correlation functions, but we first
verified that the FFT produces precisely the same results 
as slow, sliding cross-correlation.
Our speckle fingerprint images are 1024 by 1024 pixels, so they produce cross-correlation images with 2049 by 2049 points. The coherent speckle information, which is 2-3 pixels wide in the speckle patterns, is transformed into a peak 5-7 pixels wide in the correlation map.
Figure \ref{fg:corr_toon} shows a typical cross-correlation map calculated using 
the two entire speckle patterns.  Note that the shape of the
mountain under the coherence tree is shaped like a rounded
cone.  When the tree gets small---as the two speckle patterns
become almost uncorrelated---the rounded shape makes it tricky to
subtract the background properly.

To provide information about the quality of our normalized
cross-correlation values, we divided the whole speckle pattern
into 8 to 15 square regions containing 100 by 100 pixels within
which the intensity was nearly constant.  In this way we removed
the distortion produced by the shadow of the blocker, its 
support arm, and by camera defects and burns.

This piece-by-piece analysis made it much easier to separate 
the tree from the mountain. 
Because the variation in the average intensity over the 
small regions is much smaller than that over the entire image,
the mountain now decays linearly away from the tree as shown in 
Fig. \ref{fg:subtraction_toon}.
Consequently, the small regions could always be reliably 
fit with simple linear functions.
This made it much easier and more reliable to subtract the
background.

It is also very useful to note that this type of speckle analysis completely separates the coherent signal from the diffuse, incoherent scattering present in an image.  Indeed, so long as the speckle signal is identifiable and separable, then any incoherent signal is eliminated.  This calculation is capable of introducing some error, however it is usually not sufficient to detract from the values obtained for the correlation coefficients. Even when the cross-correlation becomes difficult to identify, the correlation coefficient $\rho{\rm (a,b)}$ is still normalized by the auto-correlation of each image which remain well defined and comparably large.  

\section{Experimental Results
}

As explained in the previous section, we used normalized correlation
coefficients to extract information about the correlations between
the magnetic domain configurations versus their applied-field history.

We addressed
the following questions: Are the domains precisely the same each
time we go around the major hysteresis loop?
How are the domains related at the complementary points
on the major loop?
How are the domains related for different points within the same
(ascending or descending) branch of the major loop?
The first two questions are about major loop microscopic
RPM and CPM, or inter-loop correlations. The third question
is about major loop microscopic half-loop memory (HLM) values
that probe the intra-branch correlations.  We present our answers
to these questions in turn below.

\subsection{Major Loop Microscopic Return-Point Memory and Complementary-Point Memory
}

The first question that we addressed was whether our samples exhibited
major loop microscopic RPM. To do so, we compared pairs of speckle
patterns collected at the same point on the major loop, but separated
by one or more full excursions around the major loop.
The second question that we addressed was whether our samples
exhibited major loop microscopic CPM. To do so, we compared
pairs of speckle patterns collected at one point on the
major loop with the speckle patterns collected at the
inversion-symmetric complementary point on subsequent cycles.

\begin{figure*}
\includegraphics[width=18cm]{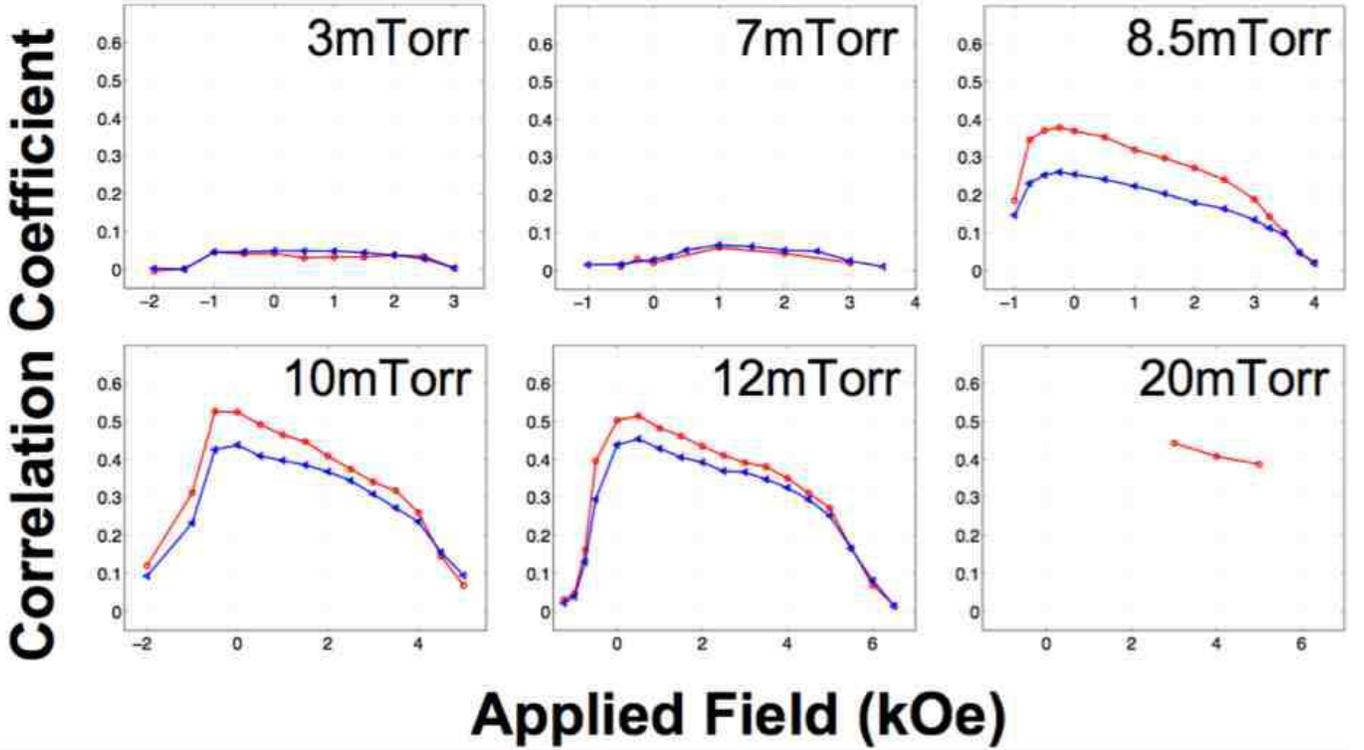}
\caption{(color online)  The measured RPM coefficients (red circles) 
and CPM coefficients (blue triangles) for all six of our  
samples along their major loop.  The 3 mTorr sample (top left) shows no  
evidence of any memory effects.  The 7 mTorr sample (top center) shows 
the possibility of an extremely small non-zero RPM and CPM.   
For each of the more disordered samples there is a sharp rise in the
RPM and CPM that correlates with the initial domain growth followed 
by growth to a maximum value and then by a slow decrease as the sample 
is taken towards reversal.
}
\label{fg:rough_rpm_cpm}
\end{figure*}

Our results for each sample are shown in Fig. \ref{fg:rough_rpm_cpm},
where the measured magnetic field dependence of the RPM and CPM
correlation coefficients for each sample are shown.
The data shown is for many excursions around each major loop.
We did not observe any RPM or CPM for the 3 mTorr sample.
The 7 mTorr sample exhibited an extremely small RPM and CPM---so
small that we could not determine whether there
was any RPM-CPM symmetry breaking.
For all of our disordered samples, 
{\it {i.e.}}, for 8.5 mTorr and above,
we measured non-zero RPM and CPM values that had their
RPM-CPM values slightly symmetry broken---our measured CPM coefficients are
consistently  a little smaller than the corresponding RPM 
coefficients. It is worth noting again that we verified with magnetometry that our samples demonstrated
perfect macroscopic return point memory.

\begin{figure}
\includegraphics[width=8cm]{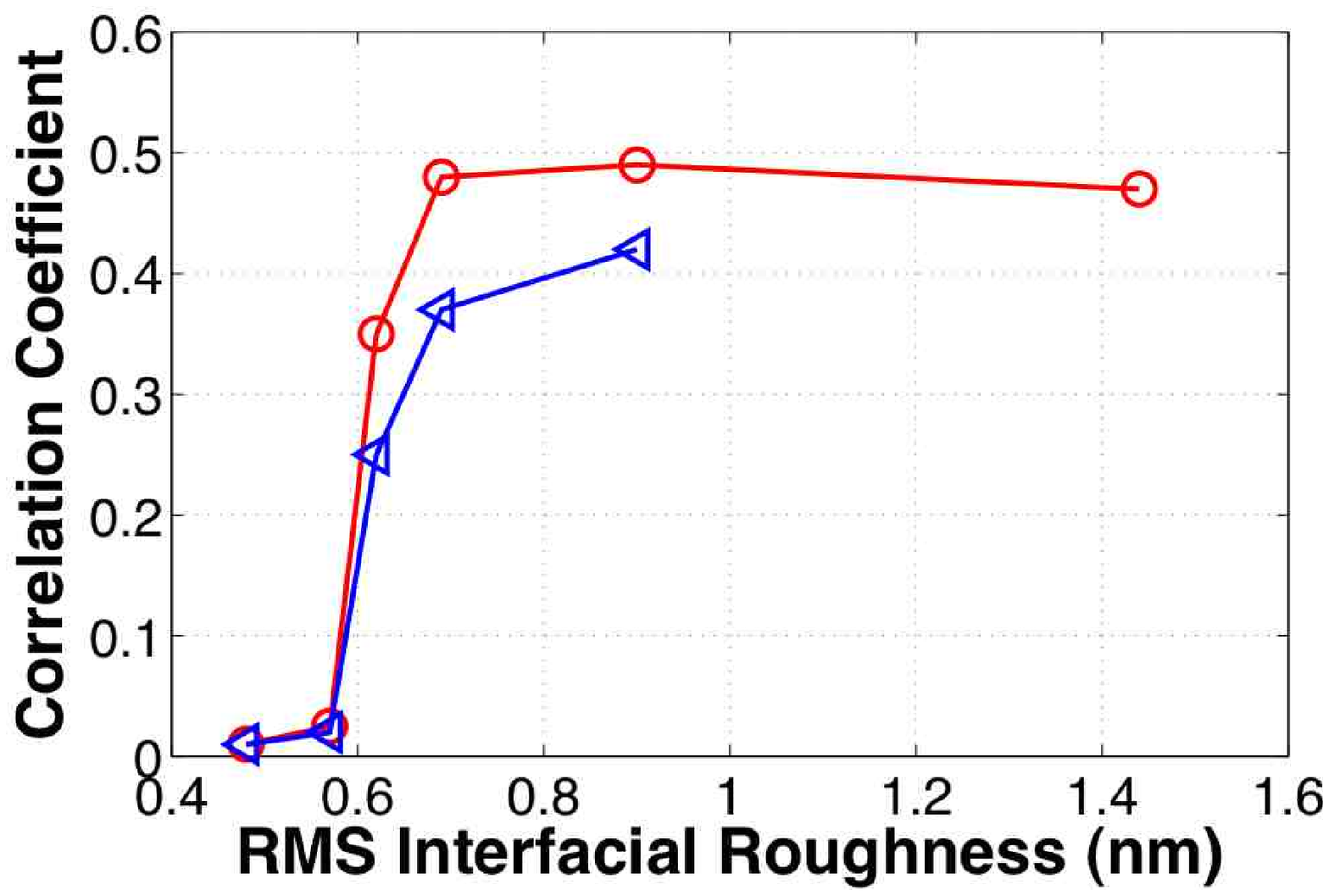}
\includegraphics[width=8cm]{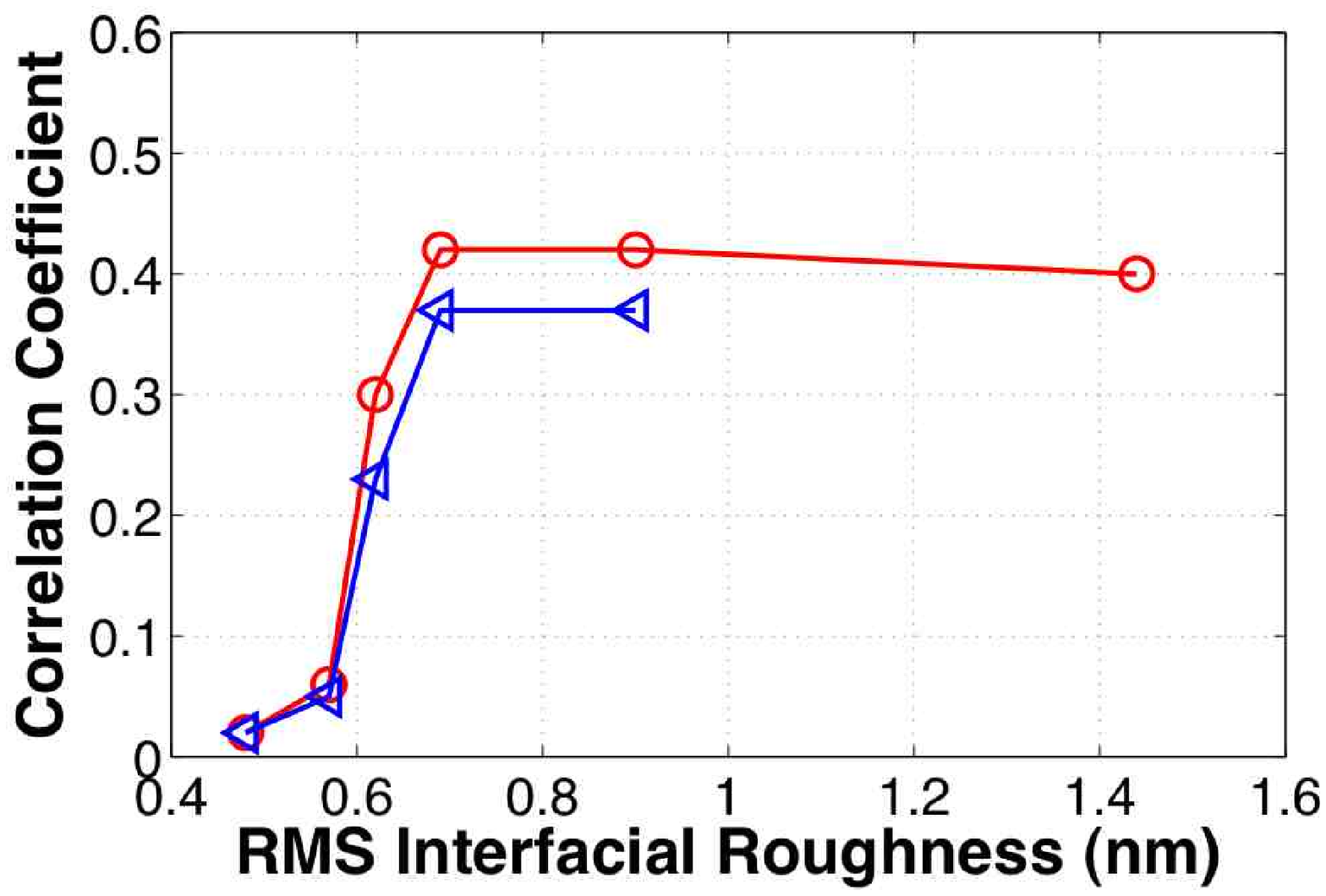}
\includegraphics[width=8cm]{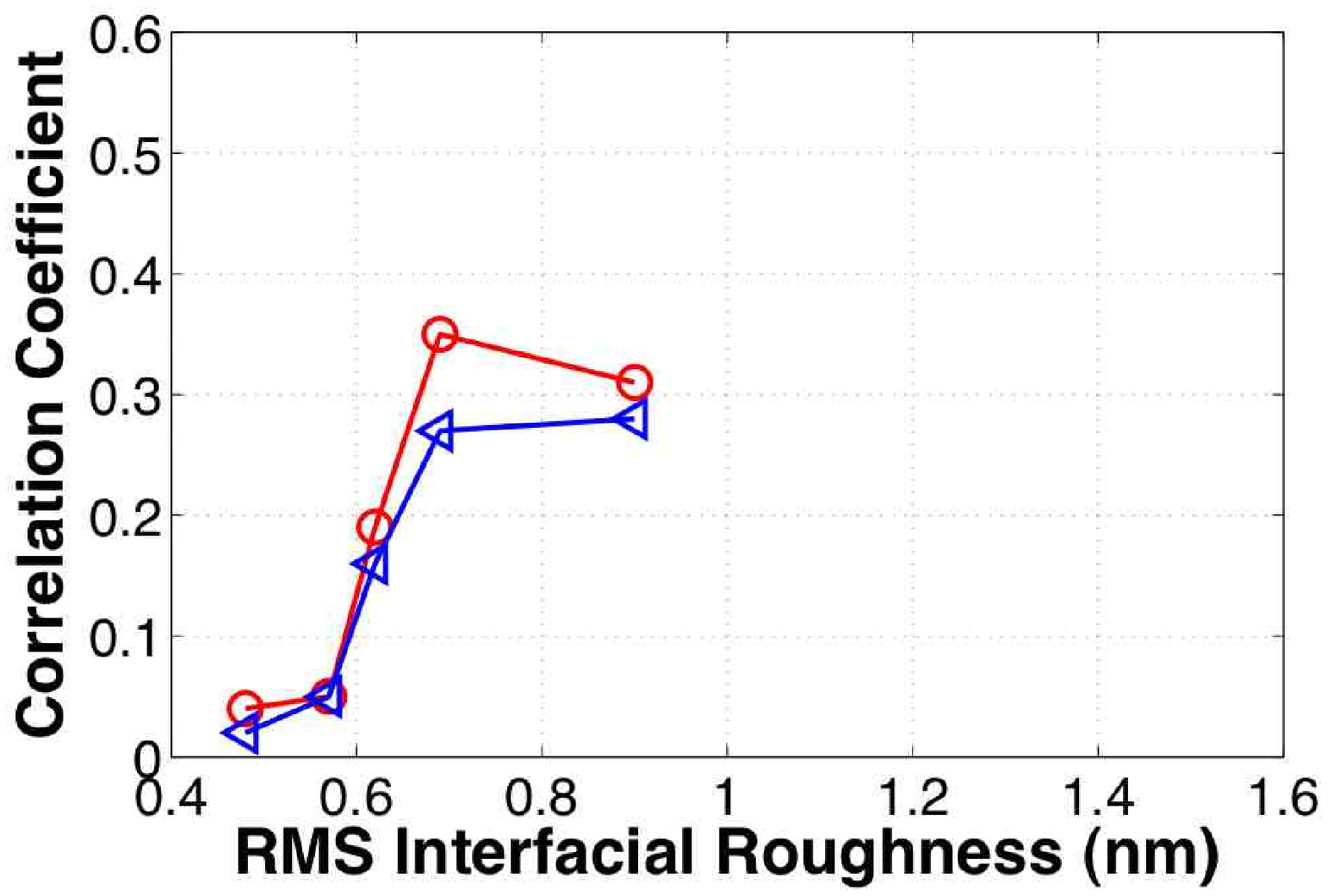}
\caption{
(color online)
The RPM values (red circles) and the CPM values (blue triangles)
measured for different sample magnetizations plotted versus the 
respective measured {\it {rms}} roughness. 
The top figure shows the results obtained for the reduced magnetization $m = -0.4$. 
The middle shows the results for $m = 0.0$ which is the coercive point. 
The bottom shows the results for $m = +0.4$.  Note that the RPM and the CPM
exhibit similar rapid growth followed by apparent saturation independent of
the value of $m$.
}
\label{fg:rpm_cpm_hc}
\end{figure}

In order to properly compare the memory of the different samples, it is helpful to plot them using the same scale.  By dividing each actual measured magnetization value, by the saturation magnetization for each sample we obtain a relative measure of the magnetization values. The value $M/M_S$ is known as the reduced magnetiztion $m$. Fig. \ref{fg:rpm_cpm_hc} shows
our major loop microscopic RPM and CPM correlation coefficients,
measured at room temperature for three values of the reduced magnetization,
$m=-0.4$, $m=0$ the coercive point, and $m=+0.4$, for each sample. 
The RPM and CPM curves are plotted versus the sample's measured {\it {rms}}
roughness.
As noted above, our smooth samples have
essentially zero RPM and CPM values.
In contrast, all of our rough samples exhibit
RPM and CPM correlation
coefficients that increase and saturate as the roughness 
increases, but never grow to unity.
This increase starts precisely where the
major loops change from being nucleation-dominated
to being disorder-dominated.

%\begin{figure}
%\includegraphics[width=8.5cm]{eb3.ps}
%\includegraphics[width=8.5cm]{eb12.ps}
%\caption{(color online) RPM and CPM values shown for the 3mT (top) and 12mT (bottom) sample with error-bars.  The fact that the values are indeed slightly larger than zero for the 3mT sample may be a real, but small correlation, or it may indicate that there is a small part of the calculation that systematically raises the $\rho$ values.  The errorbars here are much larger in the 3mT sample due to the difficulty in calculating a correlation coefficient near zero. The data proceeds from negative to positive saturation.
%}
%\label{fg:err_3mt}
%\end{figure}

In addition, our measurements establish the 
following interesting trends about the RPM and the CPM:
\begin{enumerate}
\item Neither correlation coefficient depends on the number of
intermediate loops between the correlated speckle patterns.
This indicates that the deterministic components of the
RPM and CPM are essentially stationary.
This implies that the deterministic component of the
memory in our system is largely
reset by bringing the sample to saturation.
It also strongly suggests that the same disorder is
largely producing 
at least the deterministic component of
both the RPM and the CPM.
\item The RPM correlation coefficients
are consistently a little larger than the CPM
correlation coefficients.
This demonstrates that the system acts in nearly the same way
under positive and negative magnetic fields. So this implies
that much of the disorder must have spin-reversal symmetry.
This might be produced by random anisotropy, random bonds,
or random coercivity, but not by random fields.
This also suggests that the right physics might be captured
by combining an RFIM model (which has RPM but essentially no
CPM) with any of the other microscopic models
(RAIM, RBIM, and RCIM) which all predict identical RPM and CPM.
\item The correlation coefficients are largest near
the initial domain reversal region. This suggests that the 
subsequent decorrelation is produced by the domain growth.
\item The correlation coefficients decrease monotonically
to their minimum values near complete reversal.
This suggests that the decorrelation  
is produced by the domain reversal.
\end{enumerate}

\subsection{Half-Loop Memory (HLM); The Domain Configuration Correlations 
within a Single Branch of the Major Loop}

We next studied the correlations between the magnetic domain configurations
within a single half loop---i.e., the correlations within a single ascending,
or a single descending, half loop.
We looked for evidence that the evolution of the magnetic domains 
depended upon the disorder present in the sample.
Does the level of the disorder influence how quickly or slowly the domain pattern evolves?

\begin{figure}
\includegraphics[width=8.5cm]{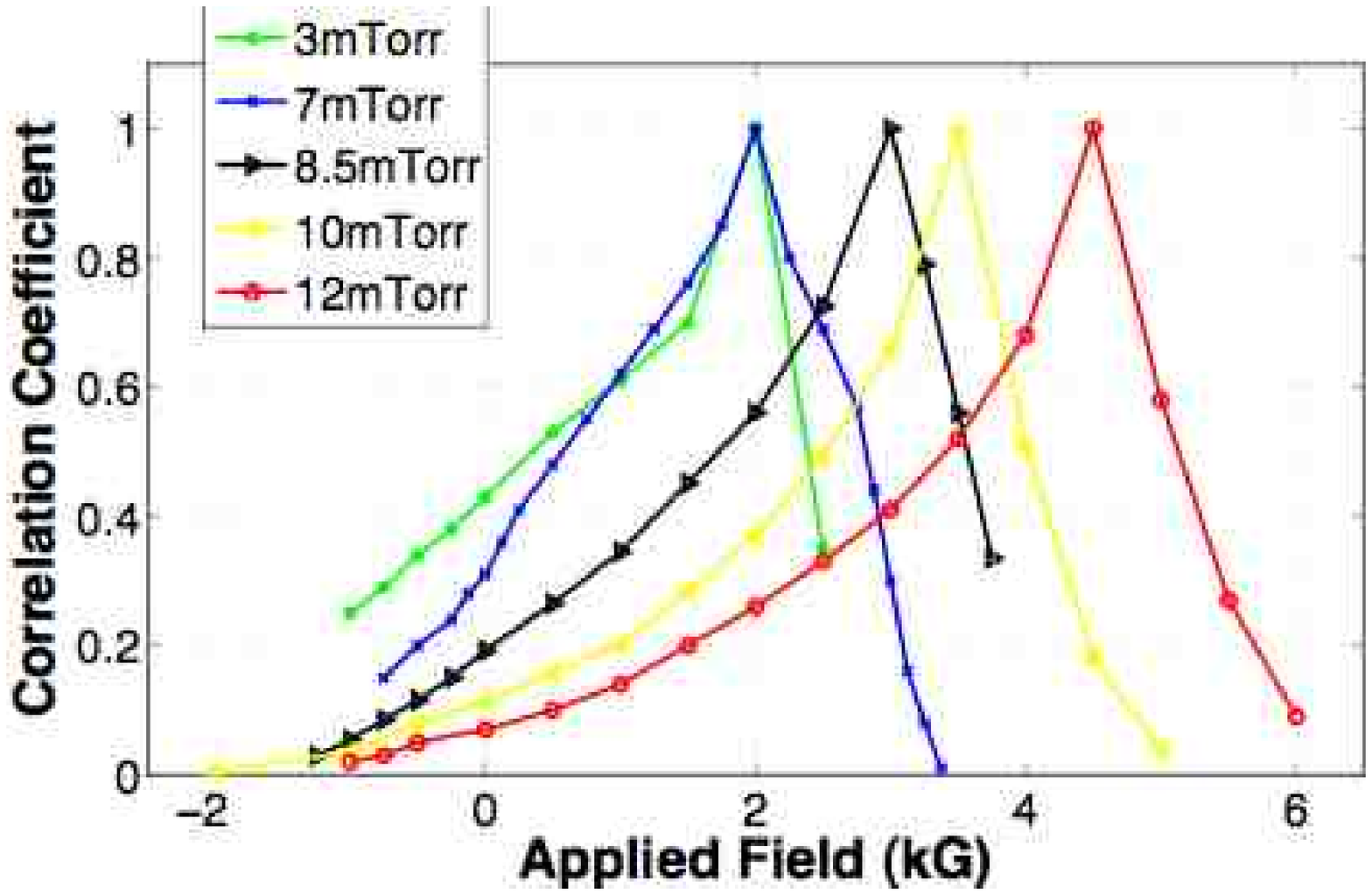}
\includegraphics[width=8.5cm]{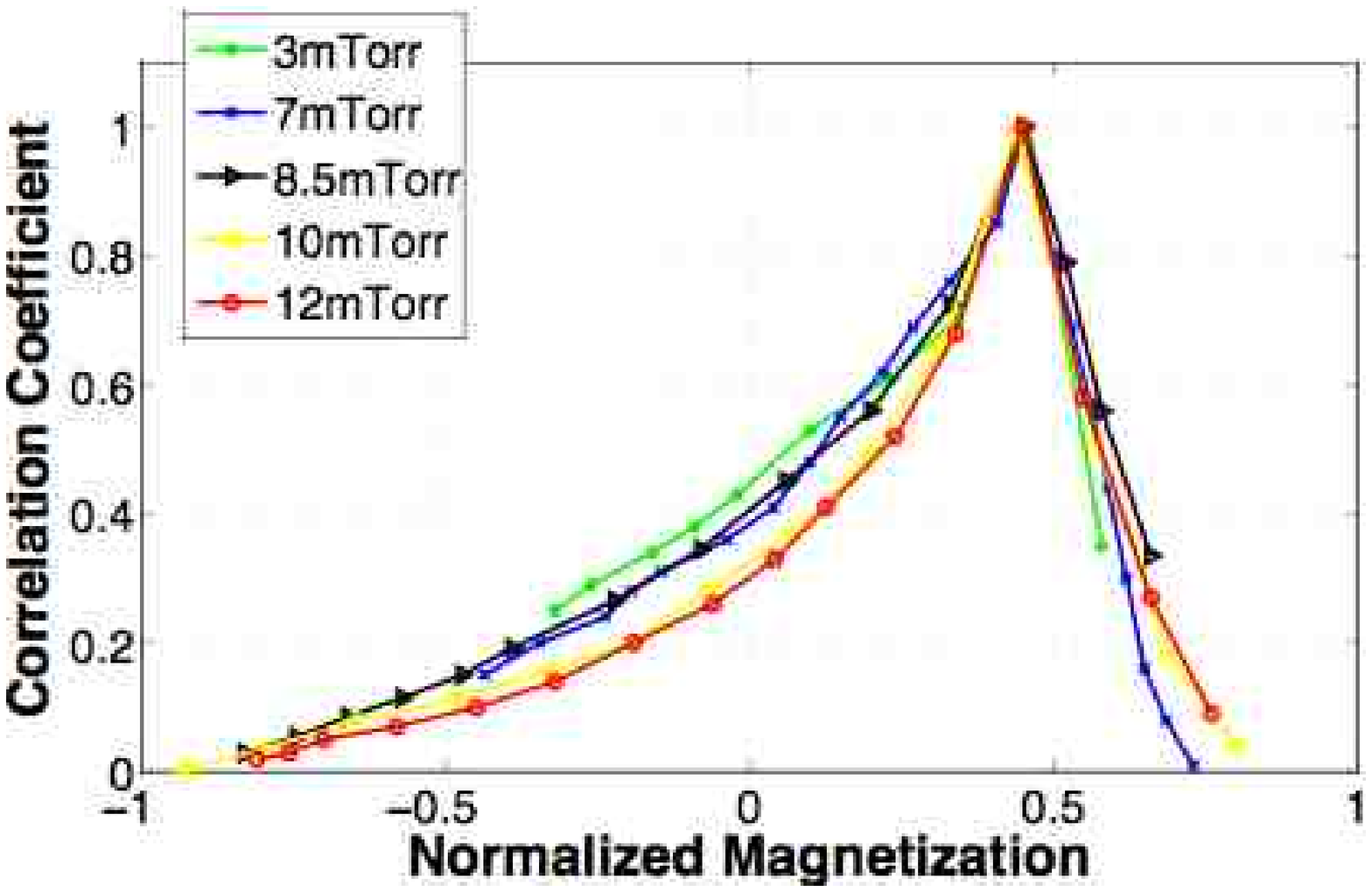}
\caption{(color online) The half loop memory (HLM) curves for $m = -0.4$
The top panel shows the HLM curves for the samples versus the applied field values.
The bottom panel shows the HLM curves versus the reduced magnetization values.  For each sample, a reference image is taken at $m = -0.4$ and then cross-correlated with all previous and subsequent images collected on the same trip along one half of the major hysteresis loop.  Thus, the degree of correlation is determined between the speckle pattern from each image with the speckle pattern obtained at $m = -0.4$ on the same trip along the major hysteresis loop.  
}
\label{fg:half_loops}
\end{figure}

To study this question, we computed the normalized cross-correlation coefficient between 
the speckle patterns taken along a single ascending or descending half-loop.
As we cycled each sample
from negative to positive saturation,
we stopped many times at fixed field values to record the speckle patterns.
By cross correlating two speckle patterns collected on the same trip along the major hysteresis loop, we are able to determine how the domain configuration is changing.

Denoting the ordered set of descending speckle patterns
by $ { 1, 2, 3, 4, ... } $, we then calculated the correlation
coefficients between all of the pairs of these speckle patterns:
\begin{itemize}
  \item $\rho(1,1), \rho(1,2), \rho(1,3),\rho(1,4),$ .... ,
  \item $\rho(2,1), \rho(2,2), \rho(2,3),\rho(2,4),$ .... ,
\item $\rho(3,1), \rho(3,2), \rho(3,3),\rho(3,4),$ .... ,
\item etc.
\end{itemize}
This allowed us to correlate each of the domain configurations
against all of the previous and subsequent measured configurations within
the same descending branch of the major loop.

When the reference pattern was taken to be the image with its reduced
magnetization $m = -0.4$,
the resulting half-loop memory (HLM) curves are shown 
versus the applied field in the top panel of Fig. \ref{fg:half_loops};
the bottom panel shows the same HLM curves plotted versus the
reduced magnetization.  Note the nice data collapse that occurs
for the plots versus $m$.

We found that the $m = -0.4$ HLM curves for our samples 
exhibit a subtle, but interesting dependance on their disorder.
Although the HLM curves in the bottom panel of Fig. \ref{fg:half_loops} 
look remarkably the similar, the low-disorder curves are
systematically above the high-disorder curves.
This indicates that the intra-loop domain
configurations in the low-disorder samples are a little
more persistent than those in the high-disorder samples.
This is consistent with the idea that the domain widths
in the low-disorder samples can expand and contract as the
applied field is changed, whereas the domains in the
high-disorder samples must break and reform.

\begin{figure}
\includegraphics[width = 7.3cm]{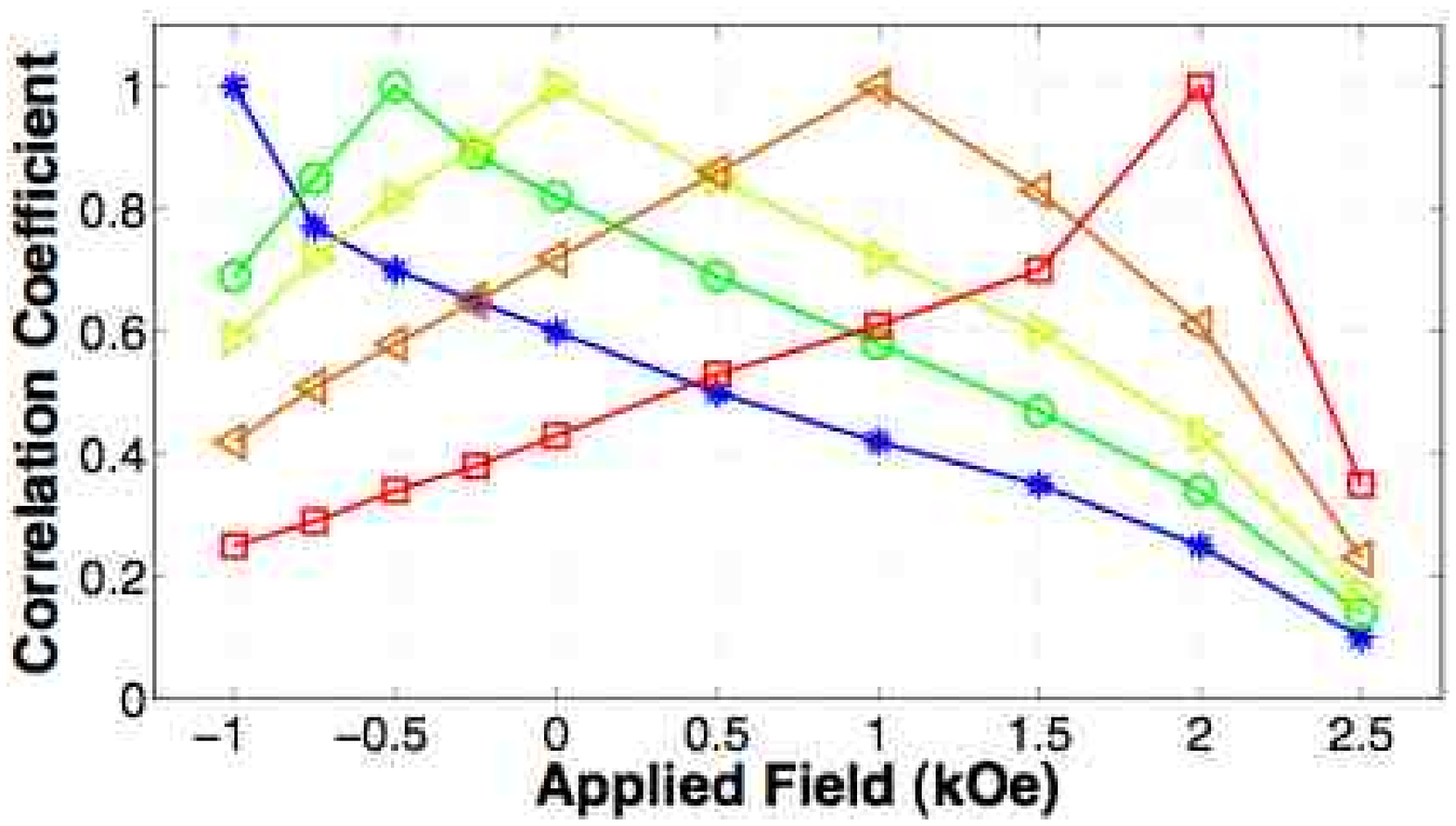}
\includegraphics[width = 7.3cm]{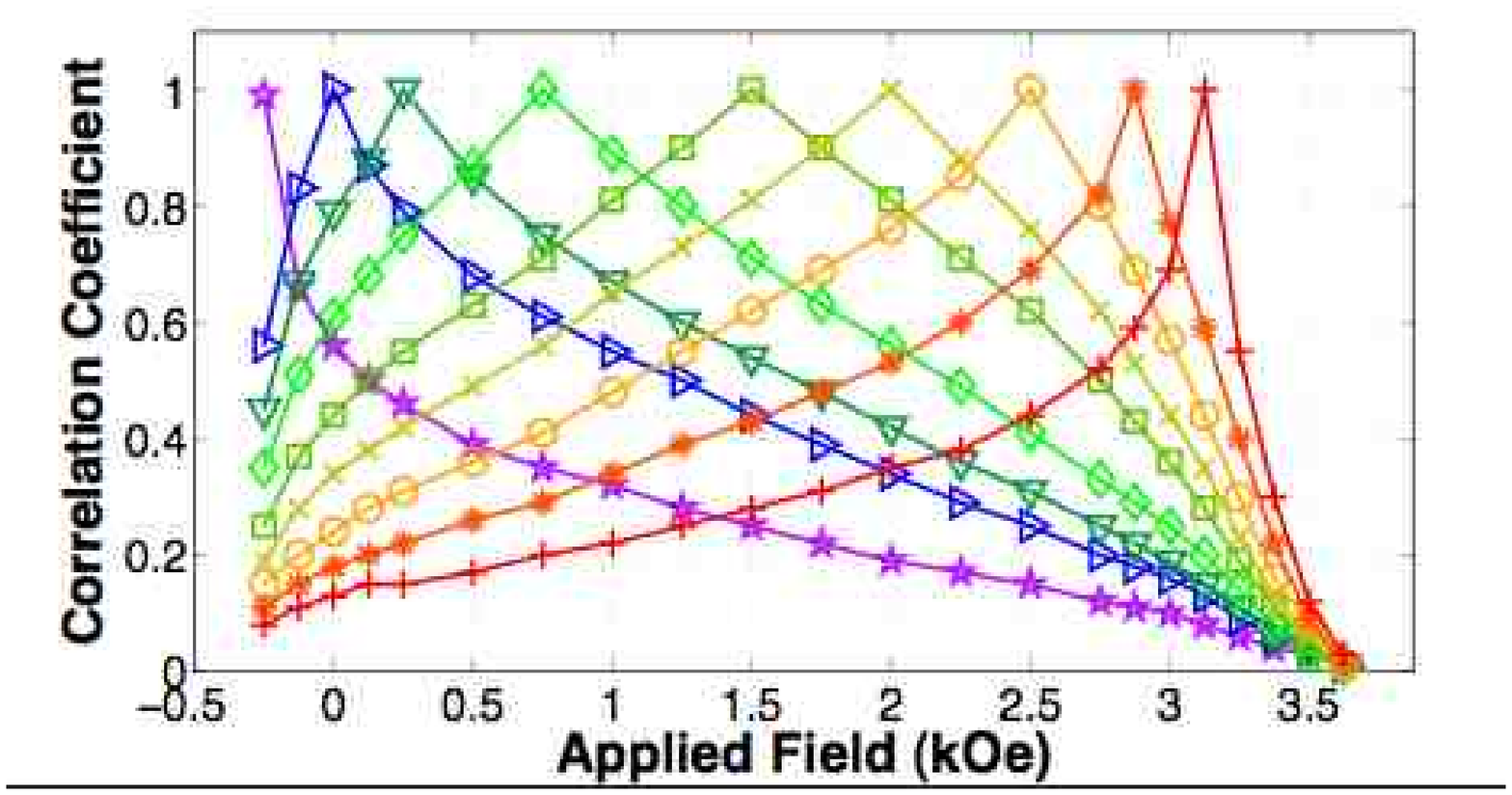}
\includegraphics[width = 7.3cm]{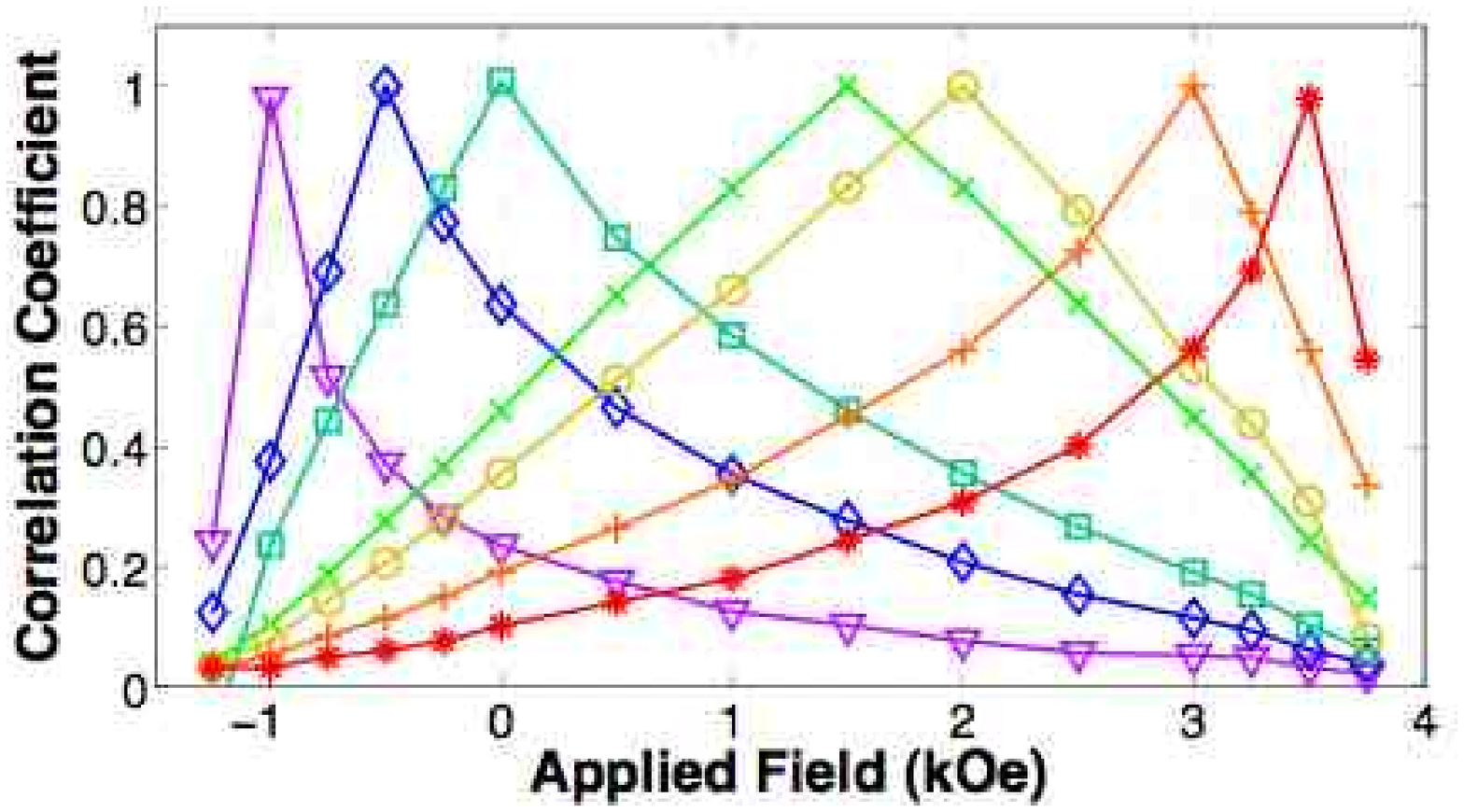}
\includegraphics[width = 7.3cm]{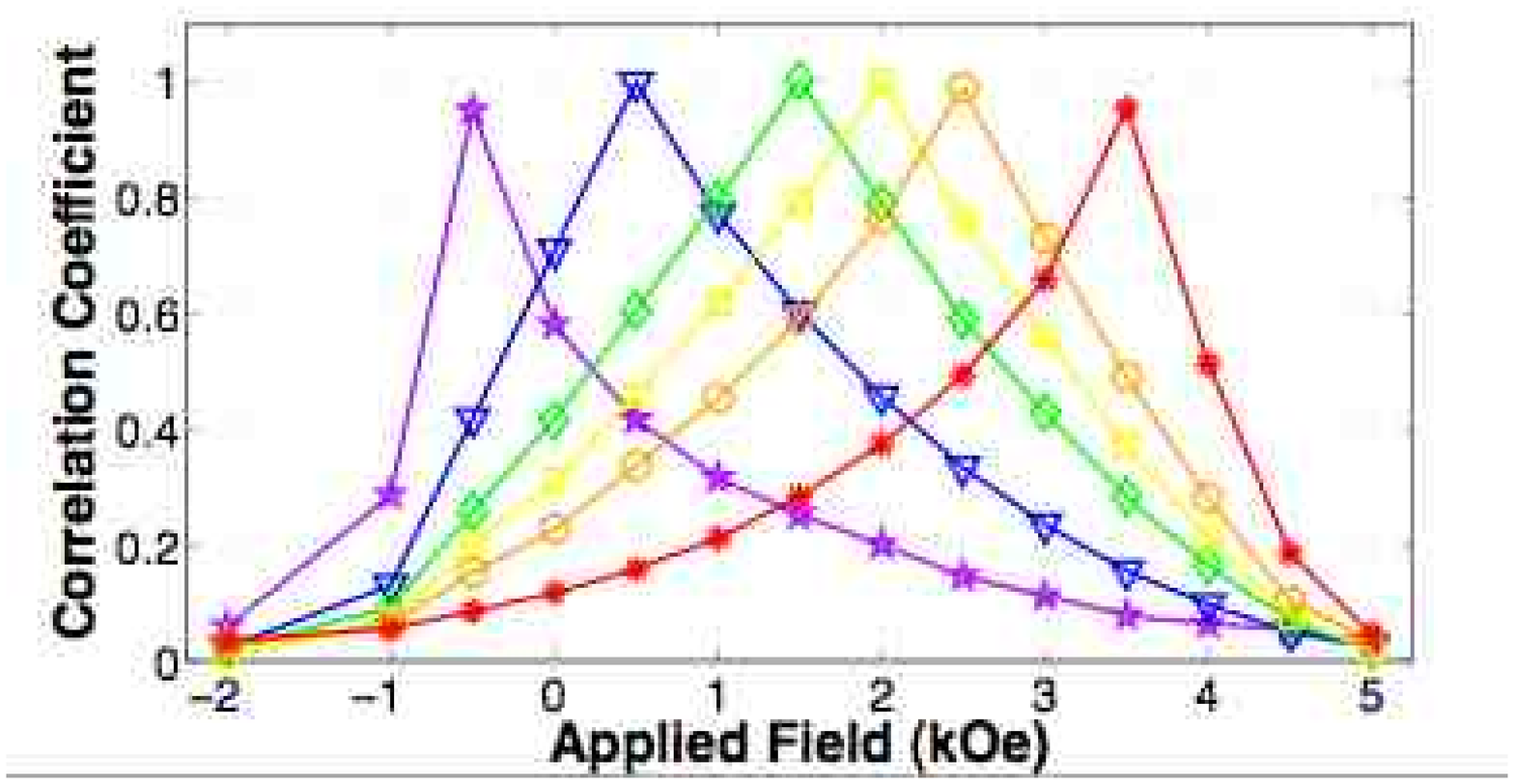}
\includegraphics[width = 7.3cm]{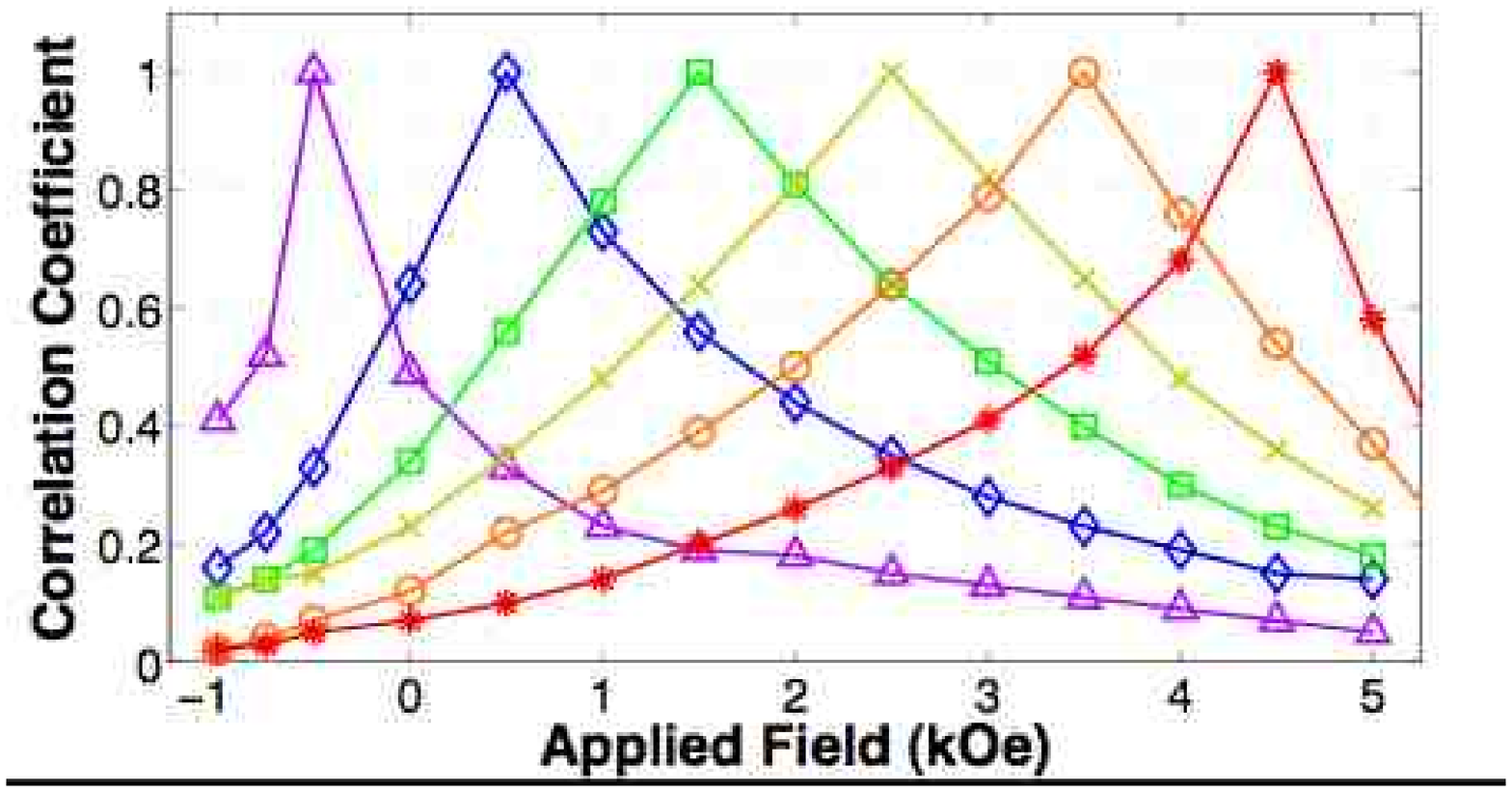}
\caption{(color online) Our measured half loop memory (HLM) curves
plotted versus the applied field values.  The HLM curves for
each sample are shown for each possible reference speckle pattern for that sample,
{\it {i.e.,}} the magnetic speckle pattern for every measured applied field value is used as 
the reference pattern for one of the curves shown.  The respective reference field
values are easily identified as the points on the HLM curves where the HLM value is unity.
From top to bottom, the panels show the 3, 7, 8.5, 10, and 12 mTorr sample HLM curves.  In each case, the magnetization of the sample was increased from negative to positive saturation. 
}
\label{all-HLM}
\end{figure}

Fig. \ref{all-HLM} shows the systematic evolution of our measured HLM curves
versus each reference image for the 3, 7, 8.5, 10, and 12 mTorr samples.
Up until now, there have been no analytic predictions or simulations for the shapes 
of, or the evolution of, these measured HLM curves.

\section{Our New Theoretical Models}

Our experimental results are the first direct measurements of the  
effects of controlled disorder on microscopic return-point  
memory and complementary-point memory.  How can we understand our  
experimental results in the light of the current microscopic disorder  
theories, and, in particular, how does nature produce the RPM-CPM  
symmetry breaking?  For spin models that are bilinear in spin  
(excluding the external field term), one might expect that the  
microscopic evolution of spins from a completely saturated state would  
lead to the same speckle patterns (or, with thermal noise or dynamical  
instabilities, to the same distribution of speckle patterns), whether one  
starts from a very large positive or very large negative applied field.  
As a consequence, the RPM and CPM correlation coefficients would be  
the same.  Our experimental results show that this is not the  
case, and we have found that this apparent experimental small
RPM-CPM symmetry breaking imposes strong constraints on any theory 
designed to describe the experiments.

In many theoretical descriptions of the magnetization process in  
systems with an easy anisotropy-axis, the local magnetization is  
described as a scalar quantity, since it is assumed that to a good  
approximation the magnetization must point along this easy axis.
Starting from this point, we can
obtain a number of different models according to the additional  
ingredients that we add.  The well known Ising model is obtained by  
coupling neighboring spins ferromagnetically.  
 
Many different microscopic disorder-induced memory systems
have been constructed by introducing different types of disorder.  
The random anisotropy Ising model (RAIM) is produced by varying the
local deviations from normal of the easy axis\cite{raim_vives}.
The random bond Ising model (RBIM) is produced by varying the
strength of the ferromagnetic coupling between spins\cite{rbim_vives}.
The random coercitivity Ising model (RCIM) is produced by varying
the coercive field that is necessary to flip each spin\cite{rcim_friedman}.
The Edwards-Anderson spin-glass model (EASG) is produced
by randomly choosing the signs of the bonds.
An additional interesting and extensively studied possibility
is to consider disorder entering in the form of a random local
field acting on each spin. In this case, the model is called the
random field Ising model (RFIM)\cite{sethna-dahmen}.  For compactness, below we will
refer the above models using the acronym RXIM.

There is an important distinction among the previous models.
The RAIM, RBIM, and RCIM are all spin inversion symmetric.
So at zero temperature, they show perfect RPM and CPM. If temperature
is included, then both the RPM and CPM become less than one,
but they remain equal. The only model that is not spin
symmetric, and can thereby explain the experimentally
observed difference between RPM and CPM is the RFIM.
The RFIM alone is too drastic because for all but extremely 
high disorder, it has essentially no CPM.  
This suggests one possible approach to explain
our results: a theory that combines the RFIM with one of
the other RXIMs possessing full CPM could produce a
system which possesses slightly symmetry broken RPM and CPM.
Another possible route to the broken RPM-CPM symmetry is via
spin-glass models.  The standard spin-glass model has 
perfect RPM and CPM at $T=0$, but when coupled to 
random fields it exhibits imperfect CPM.  Some of the relevant properties are summarized in Table \ref{tab:table2}.

We have explored the first two models, denoted Model 1 and Model 2 below,
in the most detail.  Specifically we tuned the model 
parameters to make the simulations behave as closely
as possible to our experimental results.  Both models
were able to semi-quantitatively match our experimentally 
measured behavior for:
(i) the domain configuration morphology,
(ii) the shape of the major loops,
(iii) the values of the RPM and CPM coefficients, and
(iv) the small RPM-CPM symmetry breaking,
versus both the disorder level and versus the magnetic-field-history.

We also explored one more model that was designed to produce 
the observed small RPM-CPM symmetry breaking with the minimum
physics: Model 3 combines the RFIM with a spin-glass model.
However, for this model, we have not yet tried to tune
the model parameters to match the detailed experimental behavior.
So we present it here as a minimal model that can produce a small
RPM-CPM symmetry breaking that is qualitatively similar 
to that observed in our experiments.

Model 1 combines a pure RFIM with a pure RCIM \cite{jagla_1, jagla_2}.
The essential idea was to produce a theory which combines the 
pure RFIM---with perfect zero-temperature RPM but essentially 
no CPM---with one of the other random Ising models possessing 
full CPM to thereby produce a model which possesses the 
experimentally observed slightly symmetry-broken RPM and CPM.

Model 2 explores the consequences of the type of dynamics used by the  
simulations to change the orientation of a spin \cite{ucsc_llg}.  
All of the typical RXIM  
simulations use selection and update methods which are unchanged under
a global spin flip.  The next level of sophistication
beyond simple scalar spin flips, is to use the vector Landau-Lifschitz-Gilbert 
(LLG) equation to describe the classical dynamics
of the magnetic spins. However, this equation of motion is not spin  
symmetric.  Consequently, the system will evolve differently than
it does for simple spin flip dynamics.  In fact, we found that
the major loop for Model 2 does not exhibit perfect complementary 
symmetry.

Model 3 explores the consequences of mixing a RFIM with a spin-glass  
model \cite{helmut-gergely}.
Once again, because the RFIM has RPM but essentially no CPM, 
and the spin-glass model has both RPM and CPM, this model can
be adjusted to exhibit a small RPM-CPM symmetry breaking.

The underlying symmetries and the
predicted zero- and finite-temperature behavior
of the RPM and CPM for the standard RXIMs and for the 
EASG model are shown in Table \ref{tab:table2}.
The entries in this table shows the memories for the 
zero-temperature models.
For the corresponding finite-temperature models, all of the 
``perfect" memories should be replaced by ``imperfect".  
The label ``small" for the CPM represents the fact that at 
very high disorder the RFIM will show imperfect CPM. Our three models are presented and discussed one-by-one in more detail below.

\begin{table}
\caption{\label{tab:table2}Symmetries and Memories of our 
Microscopic Disorder-Based Hysteresis Models
}
\begin{ruledtabular}
\begin{tabular}{ccccc}
Disorder & RPM & CPM & Spin-inversion & Time-reversal  \\
Model & & & symmetry & symmetry \\
\hline
RAIM        &  perfect   &  perfect     &   yes    &      yes\\
RBIM        &  perfect   &  perfect     &   yes    &      yes\\
RCIM        &  perfect   &  perfect     &   yes    &      yes\\
RFIM        &  perfect   &  small       &   no     &      yes\\
EASG        &  perfect   &  perfect     &   yes    &      yes\\

\end{tabular}
\end{ruledtabular}

\end{table}

\subsection{Model 1: The RCIM plus the RFIM}

The first model that we explored simulates
localized magnetic moments that lie in a plane and point
perpendicular to it. The local magnetization
is taken to be a scalar variable $\phi$.
We include in Model 1 the long-range dipolar interactions, the short-range  
exchange interactions,
and some
sort of quenched disorder to simulate the effect of the interfacial  
roughness.
In order to obtain more realistic results within a reasonable  
computational time,
we used a continuous variable $\phi$, instead of an Ising-like discrete  
variable.
The numerical advantages provided by a continuous variable have  
previously
been discussed in detail in \cite{jagla_1}.

\begin{figure}
\includegraphics[width=8.5cm]{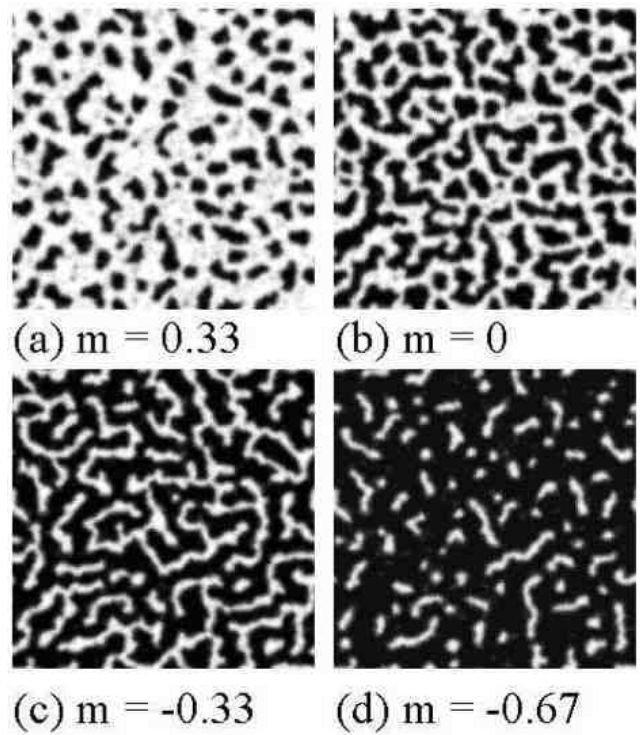}
\caption{
(color online) The simulated magnetic domain patterns for 
Model 1 showing 4 different values of $\omega_0$ at the coercive point. The simulation was started with a large applied field to saturate the magnetization in the positive direction.  Then the applied field was slowly reversed in small steps to the coercive point, allowing the domain configuration time to come to equilibrium after each step.
}
\label{fg:eduardo_domains}
\end{figure}

The total Hamiltonian for Model 1 is

\begin{eqnarray}
H=\alpha \int d{\bf r} \left (-\frac{\phi({\bf r})^2}2+\frac{\phi({\bf  
r})^4}4\right) -h\int d{\bf
r}\phi({\bf r})\nonumber\\
+\beta \int d{\bf r} \frac {\left |\nabla \phi({\bf r}) \right|^2}2
+\gamma \int d{\bf r}d{\bf r}' \phi({\bf r})\phi({\bf r}')G({\bf  
r},{\bf r}')
\label{eq:atrac}
\end{eqnarray}
For more details see \cite{jagla_1}.

The first term gives the local energy of the magnetic dipoles, favoring  
(but not
forcing) the values $\phi=\pm1$. The second term includes the effect of  
an
external magnetic field $h$, and the third term provides 
the continuum version of a local ferromagnetic interaction.
The dipolar interaction kernel $G$ is defined on a discrete numerical  
lattice by
$G({\bf r},{\bf r}') = 1/|{\bf r}-{\bf r}'|^3$
for ${\bf r}\ne{\bf r}'$, whereas $G({\bf r},{\bf r})\equiv 0$. 
It has been shown that this particular regularization adopted at
short distances does not play a crucial role in the  
results \cite{jagla_2}.

Disorder is included
through the random spatial variation of $\alpha$, namely $\alpha({\bf
r})=\bar \alpha(1 +\omega_0\eta({\bf r}))$, where $\eta({\bf r})$ is a  
random spatial
variable uniformly distributed between plus and minus one.
Consequently, the constant $\omega_0$
controls the overall strength of the disorder. This way of introducing  
disorder corresponds to that in the RCIM.  Other forms of introducing
disorder in the system (particularly random anisotropy and random
bond) should also be explored.

As anticipated above, Model 1 is spin symmetric,
and consequently
it will always predict RPM values equal to its CPM values.
To explore the possibility that
they could become different due to the existence of
some symmetry breaking field we introduced disorder
in the field through
$h({\bf r})=\bar h(1 +\omega_1\eta({\bf r}))$.
We always set $\omega_1<<\omega_0$, so that the amount
of disorder in the RFIM component was always much smaller than
that in the RCIM component. 
The time evolution of the system is obtained through
overdamped dynamics of the type

\begin{equation}
\frac{\partial\phi({\bf r})}{ \partial t}=- \frac{\delta H}{\delta \phi  
({\bf r})}
+\sqrt{2k_BT}\xi(t)
\label{1}
\end{equation}
where for convenience, time has been rescaled and
$\xi(t)$ is an uncorrelated white noise that
simulates the effects of a temperature $T$ on the system.

Through a rescaling procedure, two of the coefficients in 
the Hamiltonian
can be forced to take fixed values.  In fact, we will
assume that by appropriate time-, space-, and field-rescaling,
the
coefficients of the ferromagnetic and dipolar interactions
(namely $\beta$ and $\gamma$) have been made equal to
fixed values. For convenience in the simulations, these values
were taken to be $\beta=2$ and $\gamma=0.19$.
The new parameters on which the model depends
are now the rescaled values of $\alpha$, $h$, and $T$.
The results presented below correspond to simulations with
$\alpha=1.8$ and $k_B T = 2 \times 10^{-4}$ for
different values of the disorder set by $\omega_0$ and $\omega_1$,
and as a function of the external field $h$.

We tuned the model parameters to reproduce the experimental conditions.
We started at magnetic saturation by applying a large external field  
$h$,
so that all the local moments point in the same direction.
We then reduced the external field in small steps and obtained
the ``equilibrium configuration" by numerically solving
the Hamiltonian equation until stationarity was obtained.
Of course, the resulting ``equilibrium configuration" is actually
metastable. An example of the simulated magnetic domain configurations is shown in Fig. \ref{fg:eduardo_domains} for different applied fields.

\begin{figure}
\includegraphics[width=8.5cm]{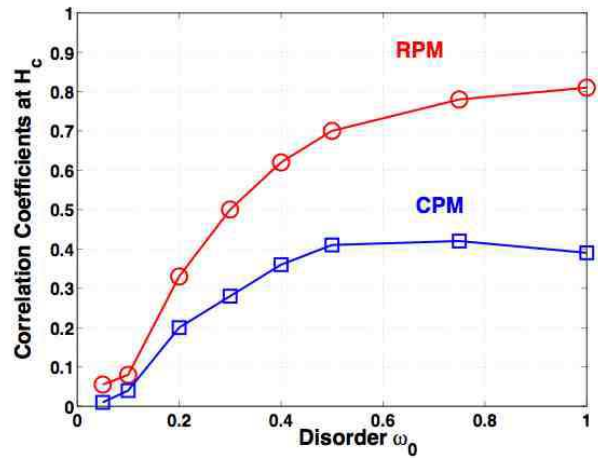}
\caption{
(color online) The calculated RPM (red circles) and CPM (blue squares) values 
versus disorder for Model 1 at finite temperature.
Note that the simulated evolution for Model 1 
is very similar to that of the experimentally measured
RPM and CPM values.  See Figure \ref{fg:rpm_cpm_hc} for comparison with experimental results.
}
\label{fg:eduardo_rpm_cpm}
\end{figure}

By using our scalar model with a ratio between the random field
component and the random coercivity component
of $\omega_1/\omega_0=0.04$, we obtained the disorder-dependent
correlation coefficients shown in Fig. \ref{fg:eduardo_rpm_cpm}.
They were calculated using the domain configurations obtained from the  
simulation.	

\subsubsection{The Effects of Combined Coercivity Disorder 
and Field Disorder on Memory}

\begin{figure*}
\includegraphics[width=8.5cm]{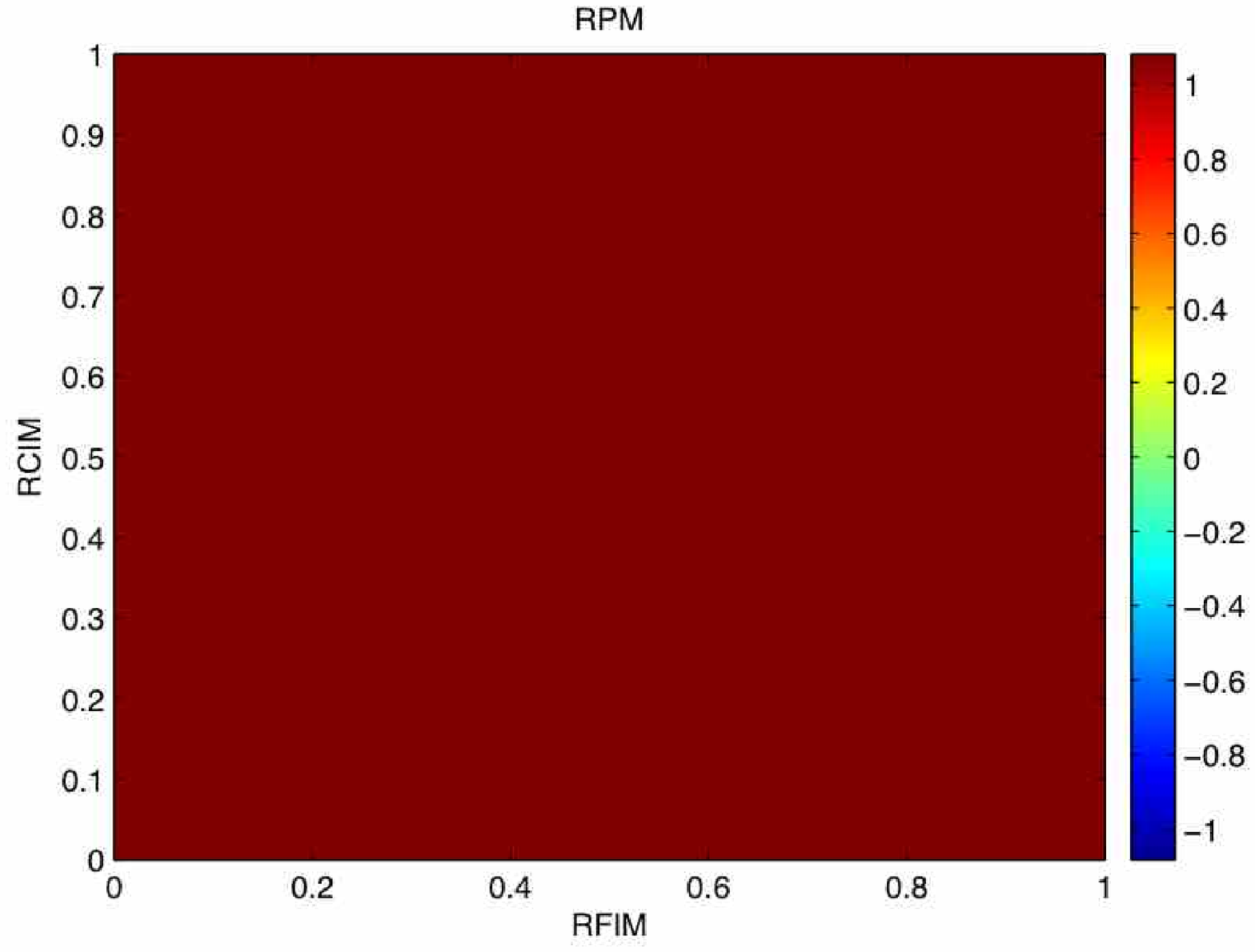}
\includegraphics[width=8.5cm]{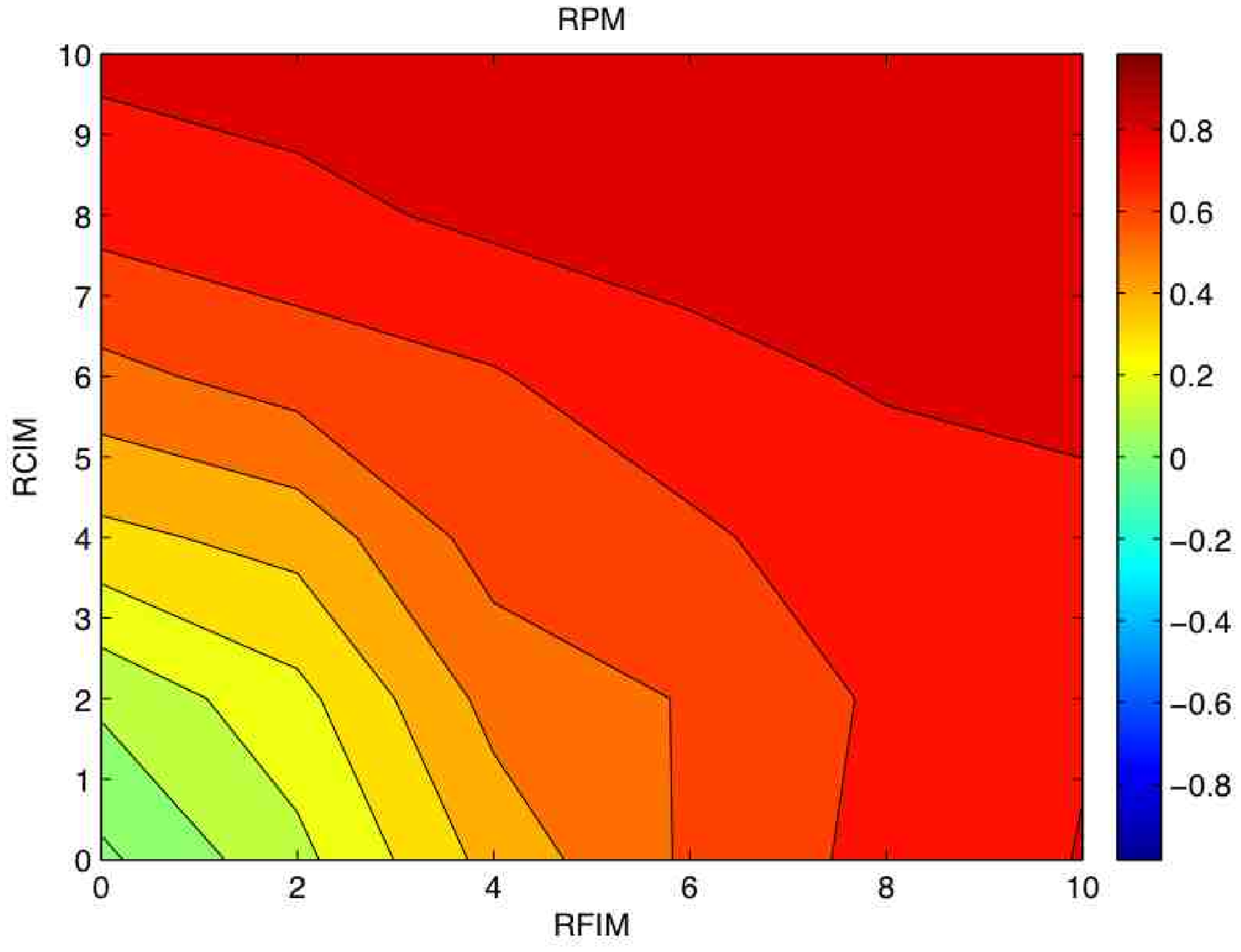}
\includegraphics[width=9cm]{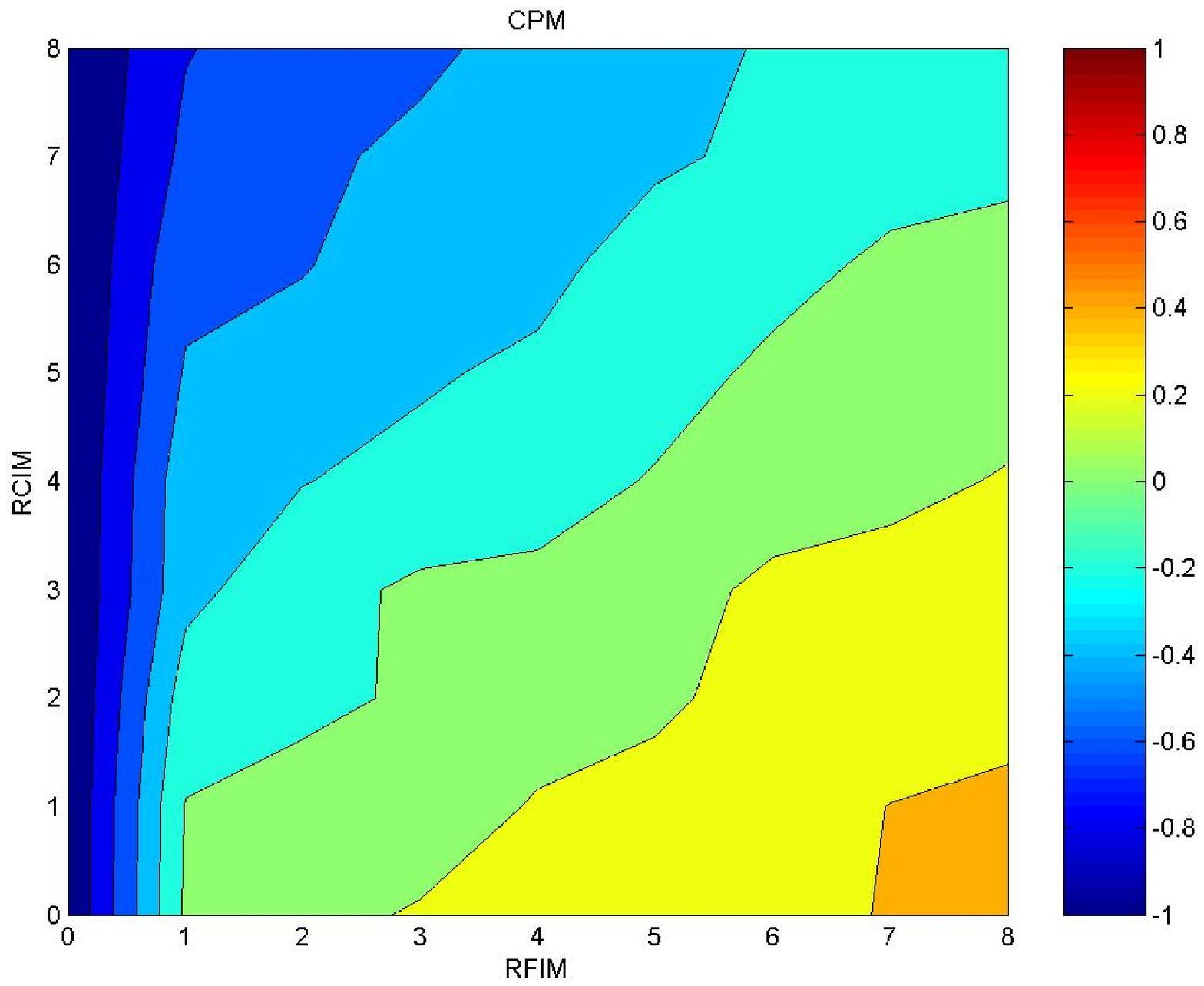}
\includegraphics[width=8.5cm]{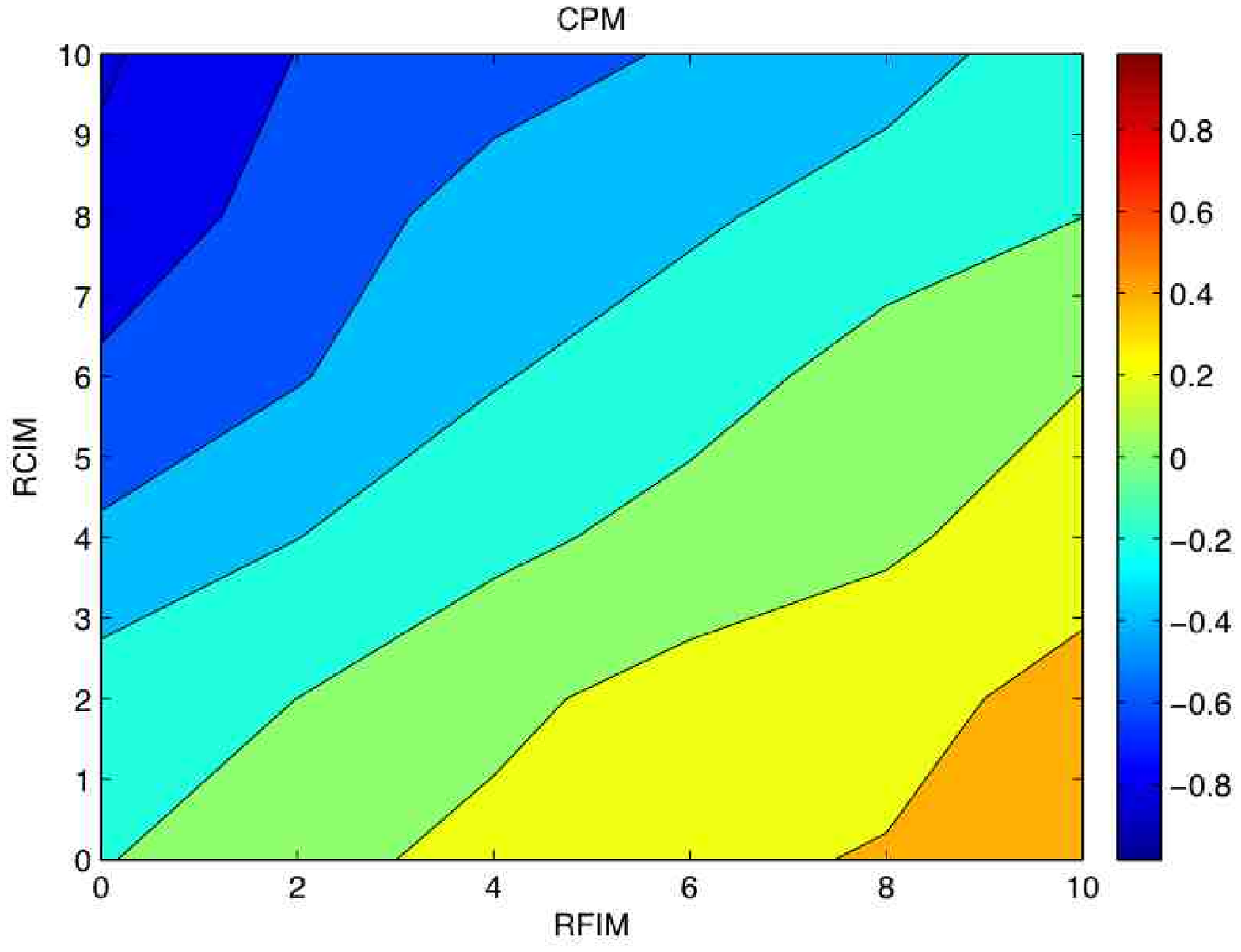}
\caption{(color online) Contour plots of the calculated RPM and CPM values for the simplified RCIM plus RFIM simulation.
The RPM values are shown on the top and the CPM values are shown on the bottom;
the $T=0$ values are shown on the left and the $T=0.1$ values are shown on the right.
The random coercivity disorder is plotted vertically
and the random field disorder is plotted horizontally.
}
\label{fg:conor}
\end{figure*}

Now that we have demonstrated that a combined RCIM plus RFIM can be
tuned to describe the resuts of our experiment quite well,
it is natural to ask about the generic behavior of this combined model.
In particular, how does its memory depend on its
two disorders---the disorder in the coercivity and the
disorder in the random field?
We used a simple discrete-spin simulation to explore
this question for both zero- and finite-temperature.
Our simulations were performed on a 64 by 64 grid of
dipolar-Ising spins. In dimensionless units, where the
near-neighbor interaction energies are unity $J=1$, our
parameters were as follows:
the random-field disorder was distributed
normally with mean 0 and standard deviation $x/3$;
the random-coercivity disorder was distributed normally
with mean 0 and standard deviation $y$.  Note that
$x$ and $y$ are the horizontal and vertical
scales in Fig. \ref{fg:conor}.

Our results for $T=0$ are shown in Fig. \ref{fg:conor}.
Note that the RPM is perfect for both the RFIM and the RCIM,
and that it is also perfect for any linear combination of the
RFIM and RCIM.
The CPM is perfect for the pure RCIM for all
values of the random coercivity disorder, but
the CPM is imperfect for the pure RFIM for all non-zero
values of the random field disorder.
The perfect CPM for both the pure Ising model
and for the pure RCIM have perfect negative CPM values
$\rho_{cpm} = -1$. Consequently the magnetic domains at the 
complementary point are perfectly anti-correlated with the magnetic
domains at the return point.

Our results for $T=0.1$ also are  shown in Fig. \ref{fg:conor}.
Note that the RPM for the pure Ising model is near zero, and that
the RPM for the pure RFIM increases as the 
random field disorder is increased---it grows from zero with
no random field disorder to about 1.0 at the
highest disorders we explored.
The RPM for the pure RCIM also increases as the 
random coercivity disorder is increased---it grows from zero with
no random field disorder to 1.0 at the
highest disorders we explored.
The CPM for the pure Ising model is near zero.
The CPM for the pure RFIM increases as the  
random field disorder is increased---it grows from zero with
no random field disorder to about +0.4 at the
highest disorders we explored.
The CPM for the pure RCIM increases as the  
random coercivity disorder is increased---it grows from zero with
no random field disorder to -1.0 at the
highest disorders we explored.  It is interesting to note that the magnitude of the correlations and anticorrelations is larger for the RCIM than the RFIM over the same range. 
The spin inversion symmetry of the RCIM pushes the system towards anti-correlation, favoring the same microscopic spin evolution on both sides of the major hysteresis loop.  In contrast, because the RFIM does not possess spin-inversion symmetry, the random field can only drive the system towards positive correlation. Both of these compete to determine the sign and magnitude of the correlation coefficients. In the range where the contributions from the RCIM and RFIM are roughly equal, the CPM is uncorrelated while the RPM increases with the magnitude of each.  

Although we have used a very simple discrete-spin simulation
done on a 64 by 64 grid of dipolar-Ising spins, note that we
obtain very similar behavior to that produced by our much more
sophisticated and realistic model presented earlier.

\subsection{Model 2: The RAIM plus LLG Dynamics}

There is another fundamental explanation for the observed RPM-CPM
asymmetry \cite{ucsc_llg}. Even when the Hamiltonian is constructed
to possess spin-inversion symmetry, the dynamics describing how
the magnetization changes do not have to be.  In fact, the dynamics
of the standard Landau-Lifschitz-Gilbert (LLG) equation break
spin-inversion symmetry

\begin{equation}
\frac{d{\bf s}}{dt} = -{\bf s} \times ({\bf B}
-\gamma {\bf s} \times {\bf B}).
\label{LLG}
\end{equation}
The first term describes the velocity with which a spin precesses
about the magnetic field.  The second term describes the damping
of the precessional motion produced as the spin aligns along the
magnetic field.

Under LLG dynamics, the spins undergo damped precessional
motion about their local
magnetic fields.  When both
the external field and all the spins are reversed,
the orientation of each spin and its precesssion are reversed. The
precessional motion of the spins ({\it {i.e.}}, their motion perpendicular
to their local fields) is not reversed, whereas the relaxational
motion (parallel to the local fields) is reversed.
As a consequence, for a disordered system,
the evolution of the magnetic domains when starting from a large
negative field is not the mirror image of the evolution starting
from a large positive field.  

Therefore the spin-inversion
symmetry of the Hamiltonian that completely determines
the equilibrium static
properties does not control the non-equilibrium dynamics
that are relevant for magnetic hysteresis.  We first
observed this effect for a vector spin model using the LLG equations
to describe a set of magnetic nanopillars \cite{ucsc_llg}.
There we found that the major hysteresis loop was not
symmetric under inversion of the applied field and
the magnetization despite the fact that the Hamiltonian
displayed this symmetry.

Model 2 that we describe below attempts to capture the physics of
the magnetic domains in our experimental CoPt layered system 
and it has many
parallels with the scalar approach described for Model 1
above. We will
assume that (1) The films are disordered on the scales relevant to
pattern formation, but are strongly anisotropic. (2) The easy axis
has small random deviations away from the direction  
perpendicular to the film.

From electron micrographs of similar sputtered films, 
we see that the layers become increasingly rough and non-planar as the
disorder is increased \cite{eric-prb}. 
Consequently, even though the physics of the perfect material
dictates a strong anisotropy, the local direction of the easy
axis will no longer be precisely perpendicular to the film. 
Because it varies randomly in space, we write the anisotropic
contribution to the Hamiltonian as

\begin{equation}
\mathcal{H}_{ani} = -\alpha \displaystyle\sum_{i}
({\bf s}_{i} \cdot {\bf \hat n}_i)^2,
\label{ani}
\end{equation}
where $\alpha$ is a model parameter. Higher-order corrections
that are even in $s_i$ are also possible, but they are
not necessary to obtain qualitative agreement with our
experiments.

Disorder is included  through random variation in the easy axis $\hat{\mathbf n}_i$ for each block of spins.  To adjust the effects of the disorder, a weighting factor $w_{ani}$ was included that controls the variation from the perpendicular axis.  For small values of $w_{ani}$ the variation is small and there is little disorder.  At larger values, the disorder is heavily weighted and has a large influence.

We have also included the short-range ferromagnetic coupling $J$.
Because we are attempting to model this system as a continuum,
we take the usual approach of minimizing effects of the grid
by writing this ferromagnetic interaction in terms of ${\bf s_k}$,
the Fourier transform of the spins

\begin{equation}
\mathcal{H}_{el} = J \displaystyle \sum_{\bf k} k^2 s_{\bf k}^2.
\label{elastic}
\end{equation}

As in the scalar case, it is also crucial to include the long-range
dipolar interaction

\begin{equation}
\mathcal{H}_{dip} = -w_{dip} \displaystyle\sum_{i,j\neq i}
\frac{3({\bf s}_i \cdot {\bf e}_{ij})({\bf e}_{ij} \cdot {\bf s}_j)
- {\bf s}_i \cdot {\bf s}_j}{r_{ij}^3},
\label{dip}
\end{equation}
where ${\bf r}_{ij}$ is the displacement vector between spins $i$
and $j$, ${\bf e}_{ij}$ is the unit vector along this direction,
and $w_{dip}$ represents the strength of the dipolar coupling.

Although this is correct for point dipoles, we are modeling blocks
of spins and must include this effect in this interaction.
In particular, the short-range behavior is smoothed out
by integration in the vertical direction ~\cite{LangerGoldstein}.
We implement this as a k-space cutoff by multiplying the dipolar
interaction in k-space by a Gaussian $\exp(-k^2d^2/2)$  where $d$
is a parameter comparable to the thickness of the sample.

Finally, we include the usual interaction with the external field

\begin{equation}
\mathcal{H}_{ext} = -B_e \displaystyle\sum_i s_i^z.
\label{ext}
\end{equation}

All of the terms in the Hamiltonian are bi-linear 
in the spins and the external field 
but, as discussed above, the LLG equations are not.
Because of the symmetry breaking produced by the LLG dynamics,
Model 2 can be tuned to produce a small RPM-CPM symmetry 
breaking similar to that measured by our experiments.

We have chosen the relaxation time to be of the
order of the precessional period near saturating fields,
that is $\gamma =1$.
Although this quantity has not been measured for our CoPt films,
it has been measured for other similar materials. 
Experiments on NiFe films
show that it is very large, approximately 100~\cite{Sandler},
If it were so large in our CoPt system, it would only accentuate 
the memory asymmetry further.
Work on CoCrPt systems~\cite{Lyberatos} has indicated that the
precessional period is comparable to the relaxation time
and if this is also true for our films, then our
$\gamma = 1$ assumption is reasonable.

Summarizing Model 2, we have coarse-grained vector
spins evolving via the LLG equations with thermal noise.
The Hamiltonian incorporates local ferromagnetic coupling,
long-range dipolar forces, disordered anisotropy modeled
by a random easy axis, and coupling to the external field.
We found a range of model parameters that produced
comparable behavior to that of our experimental samples.

\begin{figure}
\includegraphics[width=8.5cm]{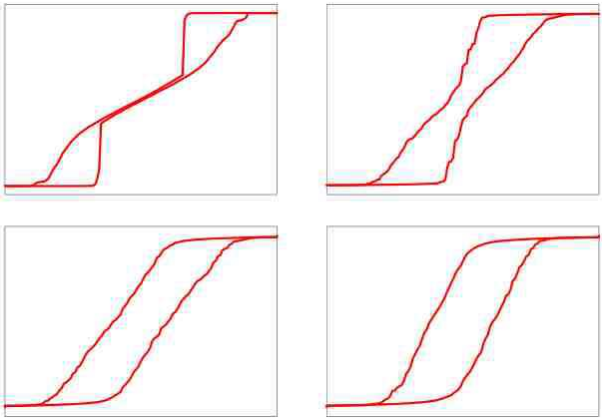}
\caption{
(color online) The simulated evolution of the major hysteresis 
loops versus the disorder for Model 2.
Note that this behavior qualitatively agrees with 
that of the experimental samples shown in Figure \ref{fg:majorloop}.
}
\label{fg:ucscloops}
%\label(fg:josh_loops}
\end{figure}

\begin{figure}
\includegraphics[width=8.5cm]{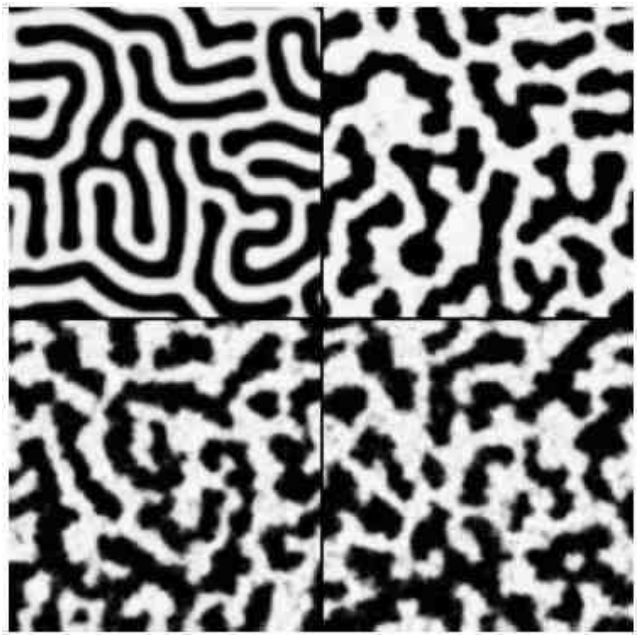}
\caption{
(color online) The simulated evolution of the domain 
configurations versus the disorder for Model 2.
Note that this behavior qualitatively agrees with 
that of the experimental samples as shown in Figure \ref{fg:6_sisters}.  The anisotropy disorder weighting parameters $w_{ani}$ are 0.001,  0.091,  0.24, and 0.33, while the dipole strengths $w_{dip}$ are 0.15, 0.105, 0.08, and 0.06 respectively.
}
\label{fg:ucscdomains}
%\label(fg:josh_domains}
\end{figure}

First, we examine the evolution of the major hysteresis loops
versus the disorder. 
The evolution of the simulated 
major loops is shown in figure \ref{fg:ucscloops}.
The disorder increases from left to right 
and from top to bottom; the simulated loop shape
for the lowest disorder is shown
in the upper left panel and
the simulated loop shape
for the highest disorder is shown
in the lower right panel.

The simulated domain configurations for different
amounts of disorder at a finite temperature 
are shown in Fig. \ref{fg:ucscdomains}.
Again the disorder increases from left to right 
and from top to bottom.
The evolution of our simulated domain configurations
versus disorder qualitatively agrees with that observed 
in the experimental samples.

For low disorder when lowering
the external field from saturation,
our simulations show that 
there is suddenly spontaneous growth of domain lines 
that fill up the system at a critical value of the field; 
this happens at constant field.
For our simulations with low disorder 
the domain morphology looks labyrinthine at remanence.
The simulated morphology and growth of the domains is
very similar to that of our experiments. 

For low disorder, the simulated hysteresis
loops look quite similar to those for the 3 mTorr samples:
corresponding to the onset of domain growth in the simulation,
there is a cliff in the hysteresis loop because the magnetization
decreases substantially during this phase of growth.
When the disorder is high, it pins the domains by destroying
translational invariance. This happens suddenly in our simulations
in a similar fashion as that in the experiments suggesting a
``critical disorder" \cite{sethna-dahmen}. 
For this ``critical disorder" and above, the spontaneous
cliff-producing growth of the domains disappears, and the domains
at remanence no longer look labyrinthine, but instead are much
more disordered.  After losing their cliff-like shape, the simulated
hysteresis loops are smooth. 
The qualitative similarities between the experimental major loops 
and the simulated major loops for Model 2 can be seen by a 
comparison of Fig.~\ref{fg:ucscloops} and Fig.~\ref{fg:majorloop}. 

\begin{figure}
\includegraphics[width=8.5cm]{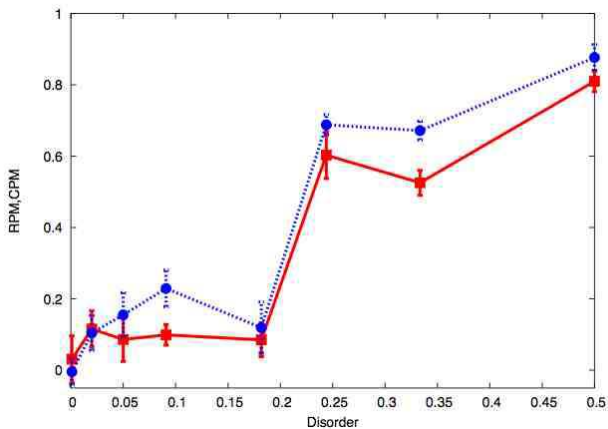}
\caption{
(color online) The RPM (red circles) and CPM (blue squares) values predicted by Model 2 plotted 
versus the disorder.
Note the qualitative agreement with the experimental measurements shown in Figure \ref{fg:rpm_cpm_hc}.
}
\label{fg:josh_rpm_cpm}
\end{figure}

We explored the memory effects in Model 2 by calculating
the correlation between pairs of domain configurations. 
Because we had direct access to the complete domain configurations, 
we calculated the correlations in real space.
The simulated RPM and CPM values for Model 2
at the coercive point as a function
of the disorder are shown in figure \ref{fg:josh_rpm_cpm}.
The RPM-CPM symmetry breaking is clearly visible.
As the temperature is increased, both curves lower in value but
the RPM curve still remains slightly above the CPM curve.
These curves clearly
grow rapidly past the ``critical disorder point" where the
simulated and experimental
loops change from cliff-like to smooth. 

The observed memory behavior in Model 2 can be explained 
as follows.
For low disorder, the spontaneous growth of the domains
is very susceptible to thermal fluctuations. If we
observe the growth of the domains for several cycles around
the major loop, we find that although the initial nucleation
points are precisely the same, the evolution past that point
is different every time. For low disorder, it appears that the
thermal fluctuations produce different domain patterns during
each cycle.
However, for high disorder, the pinning produced by the disorder
appears to constrain the domain growth and leads to significant
similarity from cycle to cycle of the domain configurations 
at the coercive point.

In summary, Model 2 simulated CoPt thin films using a spin 
symmetric Hamiltonian with LLG dynamics. Unlike in our other 
scalar models, the LLG vector {\it dynamics} is the mechanism for 
breaking the RPM and CPM symmetry. In addition to this asymmetry, 
Model 2 was also able to successfully simulate both 
the major hysteresis loops and the evolution of the domain 
configurations in qualitative agreement with our experimental 
results.

\subsection{Model 3:  A Spin-Glass Model plus the RFIM}

Motivated by the above experimental and theoretical results,
we attempted to determine the minimal model that
would capture the essential physics of the observed memory effects.
In this section, we explore the following four questions:
(1) What is the minimal model that exhibits these memory effects?
(2) Do these memory effects persist at finite temperatures?
(3) How do these memory effects depend on the disorder? and
(4) Does the RPM-CPM symmetry breaking convincingly exceed the error  
bars?
We assert that it is of general interest to study the RPM and CPM
for simple paradigmatic models, such as the
Edwards-Anderson Ising spin glass and the RFIM
\cite{sethna-dahmen, binder}

\subsubsection{Ingredient 1: The Edwards-Anderson Spin-Glass}

We start with the Hamiltonain for the Edwards-Anderson 
\cite{edwards} spin glass (EASG) given by

\begin{equation}
{\mathcal H}_{\rm EASG} =
- \sum_{\langle i, j\rangle} J_{ij} S_i S_j
- H \sum_i S_i .
\label{eq:hameasg}
\end{equation}

The spins $S_i = \pm 1$ lie on the vertices of a square lattice in two
dimensions (2D) of size $N = L^2$ with periodic boundary
conditions. The interactions $J_{ij}$ are Gaussian
distributed with zero mean and standard deviation $\sigma_J$. Our  
simulations were performed by first saturating the system by applying
a large external field $H$ and then reducing $H$ in small steps to
reverse the magnetization.

For finite temperatures, we performed a
Monte Carlo simulation and equilibrated until the average magnetization
was independent of time for each field step. For zero temperature, we
used Glauber dynamics \cite{katzgraber}
where randomly chosen unstable spins---pointing against their local field
$h_i = \sum_{\langle j\rangle} J_{ij} S_j + H$---were flipped until all
spins were stable for each field step.
These simulations converged rapidly and showed essentially
no size-dependence past $L = 20$.

We quantified the simulated RPM and CPM with $\rho_r(H)$, the overlap
in real space of the spin configuration at a field $H$ with the  
configuration at a field with the same magnitude $|H|$ after
an $n=1/2$ cycle for CPM and $n = 1$ cycle for RPM

\begin{equation}
\rho_r(H) = \frac{(-1)^{2n}}{N}\sum_{i = 1}^{N} S_i(H) S_i^{(n)}(H) \;  
.
\label{eq:q}
\end{equation}

Our results for the EASG show that this system exhibits nearly
complete RPM and CPM throughout the entire field range
for $T=0$. The strong CPM can be attributed
to the spin-inversion symmetry of the system: upon reversing all spins
$S_i$ and the magnetic field $H$, the Hamiltonian transforms into  
itself.
The observation of robust RPM and CPM answers
question 1 by establishing the EASG as a minimal model displaying
these memory effects. However, the simulated memory effects were
not perfect even at $T=0$.  This was probably due to the stochastic
nature of the updating: during each field sweep, the spins were
selected randomly for updating. Therefore, even at $T=0$, the spin  
configurations did not evolve entirely along the same valleys of
the energy landscape.  Our simulations at finite temperatures
show that the RPM and CPM decreased with increasing temperature
and remained finite, even though it would seem natural for the
thermal fluctuations to completely wash out the microscopic
memory rendering it macroscopic only.

By varying the disorder strength $\sigma_J$ in our EASG model
we showed that the RPM and CPM increase dramatically with increasing
disorder, in good agreement with the experiments. The physical
reason behind this result lies in the fact that when the disorder is  
high
the energy landscape develops a few preferable valleys and the
system evolves along these optimal valleys. This is not the case
for small disorder where several comparable shallow valleys
without a single optimal path are present.
Finally, in relation to question 4 it is noted that in the EASG the
differences between the RPM and CPM are immeasurably small.

\subsubsection{Ingredient 2: The Pure Random-Field Ising Model}

Next, we study the same memory effects in the 2D random-field
Ising model (RFIM)

\begin{equation}
{\mathcal H}_{\rm RFIM} = - J \sum_{\langle i, j\rangle} S_i S_j
- \sum_i [H+h_i] S_i \; ,
\label{eq:hamrfim}
\end{equation}
here the random fields $h_i$ were
chosen from a Gaussian distribution with zero mean and standard
deviation $\sigma_h$ ($J = 1$).
The main differences between the RFIM and the
EASG are that the RFIM does not have frustration and does not have
spin-inversion symmetry. We find that this RFIM model also shows memory  
effects
which are again stable with respect to thermal fluctuations.

Regarding question 4, the RFIM deviates from the EASG results
and correlates with the experiments: in the RFIM, the RPM and
CPM are different. The RPM is larger than the CPM for all
temperatures due to the lack of spin-inversion
symmetry in the Hamiltonian. For intermediate-to-large values of
the disorder, the CPM is negligible and, in the proximity of the  
coercive
field, the CPM correlation is even negative.
In contrast, the RPM is large in the RFIM.  In particular, for $T=0$
the RPM is perfect due to the ``no-crossing property'' of the RFIM
\cite{middleton}. Consequently,
the RPM-CPM symmetry breaking is large over much of the parameter space.  Just as for the EASG, the RFIM memory increases
due to the valleys in the energy landscape becoming more pronounced
with increasing disorder.

\subsubsection{The Combined Spin-Glass Model plus the RFIM}

Because of the way that our simulations for the pure EASG model
and for the pure RFIM do not agree with the experiments,
the following question immediately arises: Can a combined
EASG and RFIM yield results comparable to the
experiments---increasing memory with increasing disorder,
together with the RPM-CPM symmetry breaking?

In order to test this, we introduced random fields into the EASG
that act only on a small fraction (5\%) of the spins
in order to break the spin-inversion symmetry of the Hamiltonian  
[Eq.~(\ref{eq:hameasg})]:

\begin{equation}
{\mathcal H}_{\rm SG+RF} = - \sum_{\langle i, j\rangle} J_{ij} S_i S_j -
\sum_i [H + h_i] S_i \; .
\label{eq:hameasgrf}
\end{equation}
The random bonds $J_{ij}$ were chosen from a Gaussian distribution
with zero mean and standard deviation $\sigma_J$. The random fields
were chosen from a Gaussian distribution with zero mean and
standard deviation unity.

\begin{figure}
%\end{figure}
\includegraphics[width=8cm]{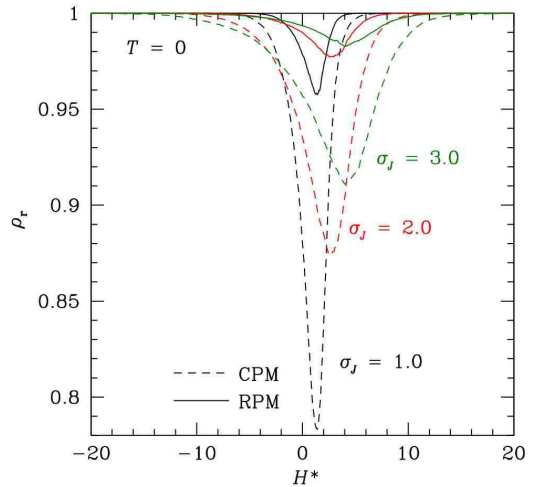}
%\hspace*{-0.55cm}
\includegraphics[width=8cm]{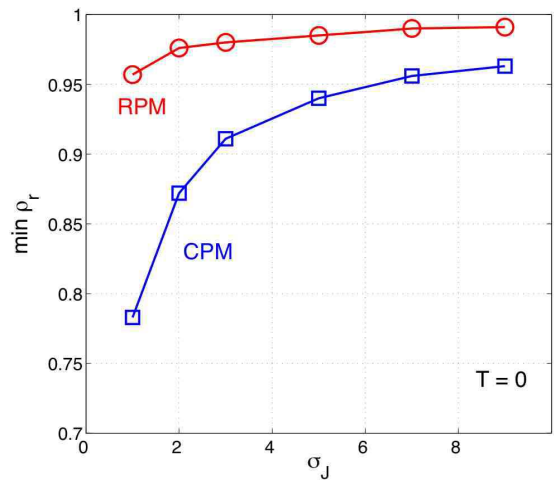}
%\end{center}
%\vspace*{-0.9cm}
\caption{(color online)
The RPM and CPM values predicted by Model 3. The field dependence for Model 3 is shown in the top panel where the predicted RPM values (solid lines) and CPM values (dashed lines) are shown for different disorder strengths. The disorder dependence for Model 3 at the coercive point is shown in the bottom panel. Note that both memory effects increase with increasing disorder and that the RPM (red circles) is always larger than the CPM (blue squares), in agreement with the experiments shown in Fig. \ref{fg:rpm_cpm_hc}. The error bars are the size of the symbols and thus have been neglected.
}
\label{fg:mixeddis}
\end{figure}

Since the dipolar interactions in our perpendicular anisotropy
films are antiferromagnetic, they introduce extensive frustration
into the system; this is the key ingredient for spin glasses.
Our experimental system also contains several possible sources
of random fields: spins frozen in by the local shape anisotropies
produced by the locally deformed environments, unusually large
crystal-field anisotropies, or by frozen-in reversed bubbles, as
reported in the same experimental system by Davies {\em et al.}
\cite{davies}.
Consequently, it is easy to imagine that all of the necessary
ingredients for the combined EASG plus RFIM are likely to be
present in our experimental samples.
Results from Model 3 are shown in Fig. \ref{fg:mixeddis} and clearly resemble the experimental results presented earlier.

\subsubsection{Summary for Model 3}

We now summarize our results for Model 3. 
We have found that our
pure EASG model, our pure RFIM model, and our combined
EASG model plus RFIM all exhibit both RPM and CPM, in which
both memory effects persist to finite temperatures and
both memory effects increase with increasing disorder.
For our pure EASG model, the simulated RPM and CPM are identical
because of the spin-inversion symmetry. For our pure RFIM, the
RPM is always much larger than the CPM because of the lack of
spin-inversion symmetry. 
Our combined ESAG and RFIM, a spin glass with diluted
random fields that break the spin-inversion symmetry,
reproduces the essential experimental results---it exhibits
both RPM and CPM, both memory effects increase
with increasing disorder, and the model can be tuned so
that the RPM is a little bit larger than the CPM.  All of these properties are present in Fig. 
\ref{fg:mixeddis}.
It will be exciting to see if experimentally realizable spin-glass systems can be shown to have these properties.

\section{Conclusions
}

Our experimental results are the first direct measurements of the  
effects of controlled microscopic disorder on the magnetic memory 
of an ensemble of magnetic domains in the such perpendicular magnetic materials.
We identified and studied 
three different aspects of the domain-level memory:
microscopic return-point memory (RPM),
microscopic complementary-point memory (CPM), and
microscopic half-loop memory (HLM).
Because our experimental observations could not be
described by any existing microscopic-disorder-based
theory, we developed new theories that do account for 
the behavior of our experimental system.  Our combined 
experimental and theoretical work sets new benchmarks 
for future work.

We found a very rich behavior of these memory properties 
in our system of perpendicular-anisotropy multilayer 
CoPt samples.
At the domain level, we found disorder-induced-partial 
RPM and CPM, and we also found a small RPM-CPM symmetry 
breaking where the RPM was consistently slightly larger 
than the CPM.

At the domain level, we found that the HLM versus the
sample magnetization was very similar for all of our
samples.  There was only a subtle effect---the low
disorder samples had slightly higher microscopic HLM
than the high disorder samples. It was surprising to us that 
the HLM effects were nearly sample independent whereas
the RPM and CPM effects were extremely sample dependent.

Our observed RPM and CPM are independent of the number 
of major loops separating the pairs of magnetic speckle 
fingerprints.  This shows that the deterministic
components of the RPM and CPM are essentially stationary, 
and implies that the deterministic memory in our system is largely reset 
by bringing the sample to saturation.  It also strongly suggests
that the same disorder is producing both the deterministic RPM and CPM.

Our measured RPM is consistently higher than our CPM.  
This slightly broken symmetry imposes severe limitations
on the possible theoretical models.  The evolution of the
RPM and CPM with disorder is also very interesting.
Both the RPM and CPM have
their largest values just after the initial domain nucleation 
takes place and then diminish
towards a minimum as saturation is approached.  
At room temperature, the memory
is imperfect---the maximum measured RPM and CPM values are 
less than one.  We suspect that this is due to the thermal
fluctuations in our samples, but have not yet demonstrated 
this experimentally.

As we increase the disorder in our samples, there is initially 
little change in the memory, then at the same level of disorder 
there is a rapid increase in both the RPM and the CPM 
followed by apparent plateaus with the RPM  
slightly larger than the CPM.
This is reminiscent of the disorder-induced transition 
predicted by the Sethna-Dahmen
RFIM work where the major loop shape changes from a gradual 
loop to a sharp loop at
a critical value of the disorder.  At the corresponding ``critical disorder 
transition" in our system, our loops also change shape, 
but more interestingly, and perhaps more importantly, both the RPM and CPM 
suddenly jump from zero to their maximum values.

Two possible explanations for RPM $>$ CPM in the disordered samples have been presented.  Within the current experimental framework, it is not possible to determine which of these two methods is a more accurate description of our system.  

 There are potential physical mechanisms that would introduce random fields into our samples.  Due to defects in the disordered samples, quenched spins may be present that do not reverse their direction even under the highest magnetic fields we were able to apply.  In effect, even though the major loop hysteresis curve appears to be constant and unchanging, there may still be spins which persist in their original direction.  These could be due to large crystal anisotropy, shape anisotropy because of the rough interfaces, or small, persistent magnetic bubbles due to incomplete saturation. In order to test for this scenario, it would be necessary to repeat the RPM and CPM measurements after saturating the samples under very high applied fields sufficient to saturate even the most stubborn spins.

In contrast, our dynamical model does not require the existence of these random fields.  Instead it requires  precessional motion to be present in the system.  We know that spins precess in the presence of a magnetic field.  However, we do not know the extent of precession relative to damping.  Measurements of the ratio of the precessional and damping terms off the LLG equation on similar systems \cite{Sandler, Lyberatos} have shown that precessional motion is quite significant.  More experiments will be necessary to determine if this is also the case for Co/Pt multilayer films.  These two suggested experiments, study under very high applied fields and determination of the precessional to damping ratio, should shed light on the true mechanism for spin inversion symmetry breaking in this system.

%We have been unable to find any previous domain-level ensemble
%studies of the RPM, CPM, and HLM in either longitudinal or perpendicular
%magnetic materials. 
There are very few direct, detailed studies of microscopic RPM, CPM, and HLM in either longitudinal or perpendicular
magnetic materials and even fewer domain-level ensemble studies of such properties. 
It will be extremely interesting
to see what our new coherent x-ray speckle metrology
technique---together with our complementary x-ray magnetic
microscopy studies---will teach us about the 
domain-level-ensemble memory in both of these 
technologically important and scientifically fascinating
magnetic memory systems.

\begin{acknowledgments}
We gratefully acknowledge the support of our work by the U.S. DOE
via DE-FG02-04ER46102, DE-FG06-86ER45275, and DE-AC03-76SF00098,
by the NSF via EAR-0216346, by the American Chemical Society via
ACS-PRF 43637-AC10, and via the Alfred P. Sloan Foundation (K.L.).
\end{acknowledgments}

\end{document}